\documentclass[fleqn,usenatbib]{mnras}

\usepackage{newtxtext,newtxmath}
\usepackage{comment}
\usepackage{amsmath}  
\usepackage{marginnote}

\usepackage[T1]{fontenc}

\DeclareRobustCommand{\VAN}[3]{#2}
\let\VANthebibliography\thebibliography
\def\thebibliography{\DeclareRobustCommand{\VAN}[3]{##3}\VANthebibliography}

\usepackage{graphicx}	
\usepackage{amsmath}	
\usepackage{xcolor}
\usepackage{multirow}
\usepackage{booktabs}
\usepackage[pagewise]{lineno}
\usepackage{rotating}
\usepackage{longtable}
\usepackage{pdflscape}

\title[Monitoring 36 Non-Repeating FRBs with FAST]{Do they repeat? Monitoring 36 non-repeating FRBs with FAST}
\author[Y. Uno et al.]{Yuri Uno$^{1}$\thanks{E-mail:yuri.uno@smail.nchu.edu.tw},
Tetsuya Hashimoto$^{1}$\thanks{E-mail: tetsuya@phys.nchu.edu.tw},
Tomotsugu Goto$^{2}$,
Shinnosuke Hisano$^{3}$,
Yi Hang Valerie Wong$^{2,4}$,
\newauthor
Arthur Chen$^{5}$,
Sujin Eie$^{6,7}$,
Simon C.-C. Ho$^{8,9,10,11}$,
James O. Chibueze$^{12,13,14}$,
Yu-Wei Lin$^{5}$,
\newauthor
Seong Jin Kim$^{2}$, 
Tzu-Yin Hsu$^{5,6}$, 
Poya Wang$^{5,6}$, 
Pei Wang$^{15, 16}$, and 
Murthadza Aznam$^{17}$
\\
$^{1}$Department of Physics, National Chung Hsing University, No. 145, Xingda Rd., South Dist., Taichung, 40227, Taiwan\\
$^{2}$Institute of Astronomy, National Tsing Hua University, No. 101, Section 2, Kuang-Fu Road, Hsinchu City 30013, Taiwan\\
$^{3}$Kumamoto University, International Research Organization for Advanced Science and Technology, Kumamoto, 860-8555, Japan\\
$^{4}$Department of Astrophysical and Planetary Sciences, University of Colorado Boulder, CO 80309, USA\\
$^{5}$Department of Physics, National Tsing Hua University, 101, Section 2. Kuang-Fu Road, Hsinchu, 30013, Taiwan (R.O.C.)\\
$^{6}$Institute of Astronomy and Astrophysics, Academia Sinica, 11F of AS/NTU Astronomy-Mathematics Building, No.1, Section 4, Roosevelt Road, \\Taipei 106216, Taiwan, R.O.C.\\
$^{7}$Mizusawa VLBI Observatory, National Astronomical Observatory of Japan, 1-2-2 Mizusawa-Hoshigaoka, Oshu, Iwate 023-0861, Japan\\
$^{8}$Research School of Astronomy and Astrophysics, The Australian National University, Canberra, ACT 2611, Australia\\
$^{9}$Centre for Astrophysics and Supercomputing, Swinburne University of Technology, P.O. Box 218, Hawthorn, VIC 3122, Australia\\
$^{10}$OzGrav: The Australian Research Council Centre of Excellence for Gravitational Wave Discovery, Hawthorn, VIC 3122, Australia\\
$^{11}$ASTRO3D: ARC Centre of Excellence for All-sky Astrophysics in 3D, ACT 2611, Australia\\
$^{12}$Department of Mathematical Sciences, University of South Africa, Cnr Christian de Wet Rd and Pioneer Avenue, Florida Park, 1709, Roodepoort, South Africa\\
$^{13}$Centre for Space Research, North-West University, Potchefstroom Campus, Private Bag X6001, Potchefstroom, South Africa, 2520\\
$^{14}$Department of Physics and Astronomy, Faculty of Physical Sciences, University of Nigeria, Carver Building, 1 University Road, Nsukka 410001, Nigeria\\ 
$^{15}$ National Astronomical Observatories, Chinese Academy of Sciences, A20 Datun Road, Chaoyang District, Beijing 100101, People's Republic of China
\\
$^{16}$Institute for Frontiers in Astronomy and Astrophysics, Beijing Normal University, Beijing 102206, People's Republic of China
\\
$^{17}$Department of Physics, Faculty of Science, Universiti Malaya, Kuala Lumpur, 50603, Malaysia
}

\date{Accepted 2025 May 28. Received 2025 April 22; in original form 2023 September 30}

\pubyear{2025}

\begin{document}
\label{firstpage}
\pagerange{\pageref{firstpage}--\pageref{lastpage}}
\maketitle

\begin{abstract}
The origin of fast radio bursts (FRBs), highly energetic, millisecond-duration radio pulses originating from beyond our galaxy, remains unknown. 
Observationally, FRBs are classified as non-repeating or repeating, however, this classification is complicated by limited observing time and sensitivity constraints, which may result in some repeating FRBs being misidentified as non-repeating.
To address this issue, we adopt both empirical and machine-learning techniques from previous studies to identify candidates that may have been misclassified.
We conducted a follow-up observations of 36 such candidates, each observed for 10 minutes using the Five-hundred-meter Aperture Spherical Telescope (FAST). 
No radio bursts exceeding a signal-to-noise ratio of 7 were detected, with a typical 7$\sigma$ fluence limit of $\sim$0.013 Jy ms.
We constrain the repetition rates of these sources using two statistical models of FRB occurrence. 
Combining our FAST non-detections with prior observations, we derive upper limits on the repetition rates of 
$\sim$10$^{-2.6}$--$10^{-0.22}$ hr$^{-1}$ under a Poisson process, and  
$\sim$10$^{-2.3}$--$10^{-0.25}$ hr$^{-1}$ 
under a Weibull process. 
This work presents one of the most stringent upper limits on FRB repetition rates to date, based on a sample size five times larger than those used in previous studies.
\end{abstract}


\begin{keywords}
(transients:) fast radio bursts -- radio continuum: transients
\end{keywords}



\section{Introduction}
Fast Radio Bursts (FRBs) are millisecond-scale, highly energetic (up to 10$^{42}$ erg) radio pulses, originating predominantly from extragalactic sources \citep[e.g.,][]{2007Sci...318..777L,Cordes2019}, with a few rare Galactic exceptions \citep[e.g.,][]{2020Natur.587...59B}. Since their discovery in 2007, the origin of FRBs has remained a profound mystery in mordern astrophysics, drawing widespread attention from the astronomical community \citep{2007Sci...318..777L}. 
Interest in this phenomenon has surged, with an increasing number of observational campaigns devoted to understanding its nature.
As of this writing, the number of FRB samples has grown to approximately 800 \citep{2023Univ....9..330X}.

FRBs are commonly classified as either non-repeaters or repeaters, depending on whether multiple bursts are observed from the same source. Non-repeaters are identified by the detection of a single burst, with no additional bursts observed despite follow-up efforts, while repeaters have produced multiple bursts. Currently, around 60 repeating sources have been confirmed \footnote{https://www.chime-frb.ca/repeaters}. 

The separation between repeaters and non-repeaters is not as straightforward as it appears. Many non-repeaters could, in fact, be repeaters whose subsequent bursts were missed due to limited observing time or the sensitivity thresholds of current telescopes  \citep{2019MNRAS.484.5500C, 2018ApJ...854L..12P, 2020MNRAS.498.1973L, 2018MNRAS.481.2320L, 2020MNRAS.494..665L}.
Ideally, extensive follow-up observations are necessary to establish an FRB's repeatability status, but such efforts are resource-intensive.  To address this challenge, \citet{Hashimoto2020} and \citet{Chen2022} proposed alternative classification frameworks based on the statistical analysis of the observed FRB properties.

\citet{Hashimoto2020} analyzed the time-integrated luminosity-duration relation $L_{\rm \nu}$--$w_{\rm int}$ and luminosity functions of FRBs, identifying a clear separation between repeaters and non-repeaters. Namely, repeaters tend to exhibit fainter luminosities and longer durations than non-repeaters (see also, Fig. \ref{fig1}). 

\citet{Chen2022}, meanwhile, introduced a classification scheme using the unsupervised machine learning algorithm Uniform Manifold Approximation and Projection (UMAP). They demonstrated that UMAP can successfully identify repeating FRBs without prior knowledge of their repeatability, and can highlight candidates potentially misclassified as non-repeaters.

This study aimes to evaluate the classification schemes proposed by \citet{Hashimoto2020} and \citet{Chen2022} using the Five-hundred-meter Aperture Spherical Telescope \citep[FAST,][]{Nan2011}, one of the most sensitive single-dish radio telescopes in the world. FAST's sensitivity surpasses that of other major radio telescopes, including the Canadian Hydrogen Intensity Mapping Experiment (CHIME), Murriyang, the Parkes radio telescope (Parkes), the Upgrade of the Molonglo Observatory Synthesis Telescope (UTMOST), and the Green Bank Telescope (GBT), by more than an order of magnitude. This enhanced sensitivity significantly improves the likelihood of detecting repeating events, which are typically an order of magnitude fainter than single bursts \cite[e.g.,][]{Hashimoto2020}.

In this paper, we report upper limits on FRB repetition rates, derived from the non-detection of repeat bursts in  follow-up observations conducted with FAST.

Section \ref{sec:target_selection} describes the criteria and methods used for target selection.  Section \ref{sec:exptime} outlines the optimization of exposure time under the constraint of a 12-hour observation window. Section \ref{sec:observation} details the FAST observing campaign and data analysis procedures. Results are presented in Section \ref{sec:result}, where we report the non-detection of repeat bursts and establish statistical upper limits on their repetition rates. Finally, Section \ref{sec:discussion} discusses the implications of our findings, and Section \ref{sec:conclusion} summarizes our conclusions.

\section{Target selection}
\label{sec:target_selection}
We selected our targets from FRBCAT \citep{Petroff2016} and the first CHIME~FRB catalogue \citep{CHIME2021_first}.

\subsection{Empirical method}
We employed the physical parameters derived by \citet{Hashimoto2020c},  including rest-frame intrinsic durations $w_{\rm int}$ and energy densities $L_{\nu}$, to identify targets from FRBCAT as of 24 Feb. 2020.
The intrinsic duration $w_{\rm int}$ is calculated by correcting the observed duration for instrumental and redshift-related  broadening effects\citep{Hashimoto2019, Hashimoto2020, Hashimoto2020c}. 
Redshifts were estimated from the observed dispersion measures (DMs), unless spectroscopic redshifts were available \citep[see][for details]{Hashimoto2022}.
$L_{\nu}$ was computed from observed fluence values, integrated over the estimated distance to each FRB.
We select non-repeating FRBs from FRBCAT based on the following criteria:
\begin{enumerate}
\item[(A1)] $\log L_{\nu}< 32$ 
\item[(A2)] $\log w_{\rm int} > 0.22(\log L_{\nu} - 32.5) + 0.33$
\item[(A3)] $-14 < {\rm Dec} < 66$,
\end{enumerate}
where $L_{\nu}$, $w_{\rm int}$, and Declination are in units of erg Hz$^{-1}$, ms, and deg, respectively.

Criteria (A1) and (A2) were empirically designed to select non-repeating FRBs that lie close to the repeater population in the $L_\nu$-$w_{\rm int}$ parameter space. 
Criterion (A3) was determined based on  the visibility constraints of FAST.

\begin{figure}
	\includegraphics[width=0.45\textwidth]{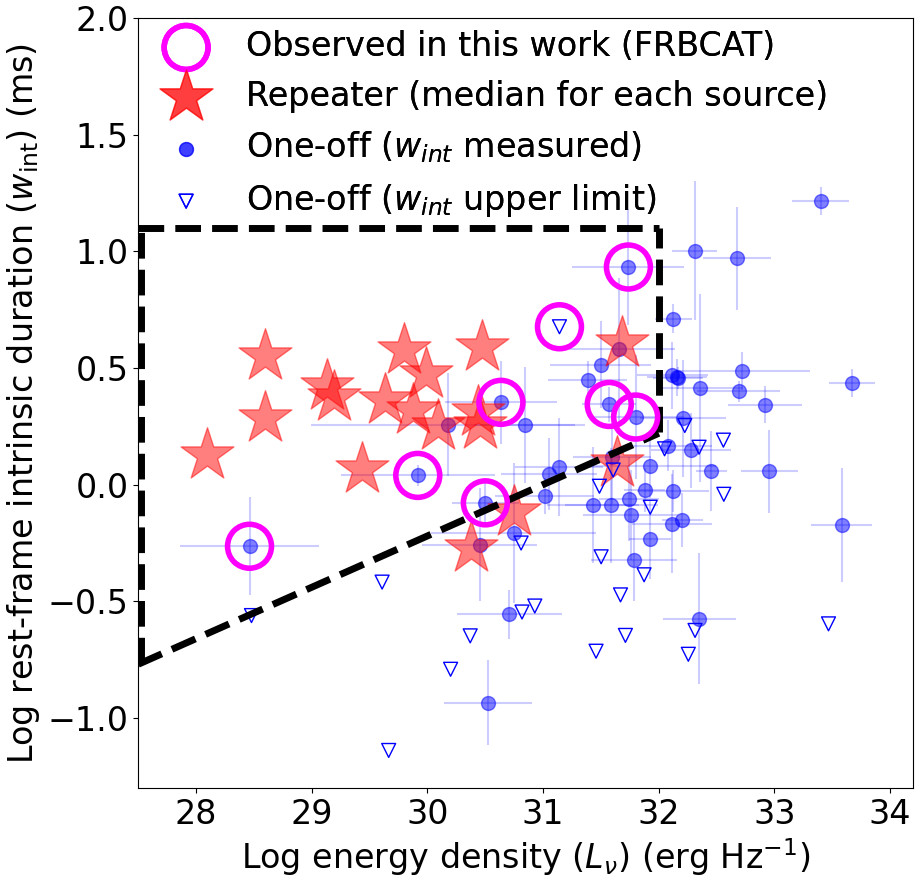}
    \caption{
    The rest-frame intrinsic duration ($w_{\rm int}$) of FRBCAT samples \citep{Petroff2016, Hashimoto2020c} as a function of the energy density ($L_{\nu}$).
    Repeating and non-repeating FRBs are shown by red and blue markers, respectively.
    Because each repeating FRB source has multiple detections of FRBs, we use the median value to present each repeating FRB source in this diagram.
    The blue circles and open triangles indicate the measured and upper limit on $w_{\rm int}$, respectively.
    The region enclosed by dashed black lines indicates our empirical selection criteria A1 and A2.
    The magenta open circles mark observed sources with FAST in this work.
    }
    \label{fig1}
\end{figure}

Using these criteria, we identified nine FRBs from FRBCAT as potential targets.
Of these, eight were observec with FAST in this study (see Section~\ref{sec:observation} for details). 
These eight targets are marked with magenta circles in Fig. \ref{fig1}, and are hereafter referred to as \lq FRBCAT\_empirical\rq sample. 
FRBs detected by the Pushchino radio telescope were excluded due to their systematically lower signal-to-noise (S/N) ratios compared to those from other telescopes\citep{Fedorova2019}.

\begin{figure}
	\includegraphics[width=0.45\textwidth]{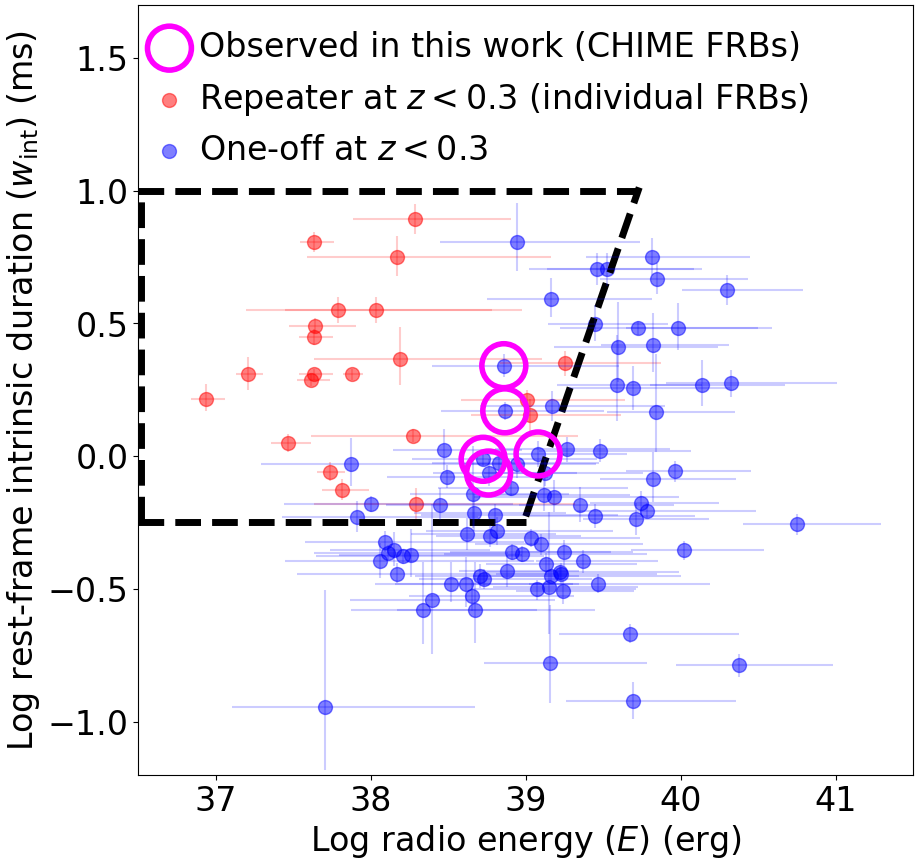}
    \caption{
    The rest-frame intrinsic duration ($w_{\rm int}$) of CHIME~FRB samples \citep{CHIME2021_first, Hashimoto2022} as a function of the radio energy ($E$).
    Repeating and non-repeating FRBs at $z<0.3$ are shown by red and blue markers, respectively.
    The individual redshifts of FRB sources are derived from observed DMs unless a spectroscopic redshift is available \citep[see][for details]{Hashimoto2022}. 
    Individual repeating FRBs are presented in this diagram.
    The region enclosed by dashed black lines indicates our empirical selection criteria B1 and B2.
    The magenta open circles mark observed sources with FAST in this work.
    }
    \label{fig2}
\end{figure}

For CHIME~FRB samples, we employ two distinct methodologies to identify target objects. 
The first methodology aligns closely with the one delineated above for the FRBCAT sample. 
In accordance with \citet{Hashimoto2022} and \citet{Kim2022}, 
we applied an empirical selection, $w_{\rm int}$ as a function of energy ($E$), to the CHIME~FRB samples.
We select the samples using the following criteria:

\begin{enumerate}
\item[(B1)] $\log w_{\rm int} > -0.25$ 
\item[(B2)] $\log w_{\rm int} > 1.7 (\log E - 66.5) + 0.25$
\item[(B3)] $-14 < {\rm Dec} < 66$
\item[(B4)] $z$ $< 0.3$
\item[(B5)] $|b| > 30$,
\end{enumerate}
where, $E$ is measured in erg units, $z$ is redshift and $b$ represents Galactic latitude, measured in degrees. 
We incorporate criterion (B4) into the CHIME~FRB samples because the disparate distributions of repeating and non-repeating FRBs in the $E$-$w_{\rm int}$ space are reported for CHIME~FRBs at $z<0.3$ \citep{Kim2022}.
We incorporate criterion (B5) to mitigate interference from the elevated radio background originating from the Galactic plane. 
Applying these criteria yields six CHIME~FRBs. 
In this study, we conducted FAST observations for five of the six identified FRBs (refer to Section \ref{sec:observation} for additional details). 
The five observed FRBs are denoted by magenta circles in Fig. \ref{fig2}. 
Subsequently, these samples will be referred to as ‘CHIME\_empirical’.

\subsection{Machine learning method}
Another approach applied to the CHIME~FRB samples is unsupervised machine learning. 
\citet{Chen2022} implemented UMAP on the CHIME~FRB samples to distinguish between repeater and non-repeater clusters within the high-dimensional FRB dataset. 
Their findings indicate that 188 non-repeating FRBs are located at clustering regions of confirmed repeaters on the projected UMAP plane \citep{Chen2022}.  Although currently unobserved to repeat, these non-repeaters could potentially be authentic repeaters. In this study, we chose 69 repeating FRB candidates as our target samples by applying criteria (B3) and (B5) to the identified 188 potential repeaters. We conducted observations for 25 out of the 69 targets in this work. 
These samples are henceforth designated as ‘CHIME\_ML’. 
Note that two CHIME sources overlap between CHIME\_empirical and CHIME\_ML.

\section{Optimised exposure time on each source}
\label{sec:exptime}
Prior to executing our FAST observations, we conduct preliminary Monte Carlo simulations to optimise the exposure time allocated to each source. We operate under the assumption that the chosen samples represent actively repeating FRB sources, each characterized by a specific burst rate, period, and active duration. FRB~20121102A and FRB~20201124A are utilised as case studies within the Monte Carlo simulations. FRB~20121102A is a well-known active repeating FRB source, with a period of approximately 156.1 days and an active phase lasting about 99 days \citep{ATel13959}. This active phase corresponds to a duty cycle of 63.6\% \citep{ATel13959}, indicting the proportion of time it remains active within a single period.

Previous FAST observations 
of FRB~20121102A 
detected 1652 bursts over a 59.5-hour exposure, corresponding to a burst rate of 28 FRBs hr$^{-1}$ during its active phase \citep{Li2021}. Using these observed parameters, we simulate observations with specific time windows, randomising the observation time within a single period. 
The total observation time includes the exposure time and a ten-minute overhead for each source.

If the simulated observation falls within the active window of a repeating FRB source, the expected number of FRB detections is calculated as the product of the assumed FRB rate and the exposure time within the active window. 
Shorter exposure times per source allow for observing a greater number of samples since the total observation time is fixed at twelve hours.
In contrast, longer exposure times per source reduce the number of samples observed. 
A single set of the simulated twelve-hour observations includes exposures on multiple sources and overheads.
We repeat the set of observations 10,000 times for a given exposure time on each source. 
The expected numbers of repeater confirmations are estimated as a function of the exposure time on each source.
\begin{figure*}
	\includegraphics[width=0.45\textwidth]{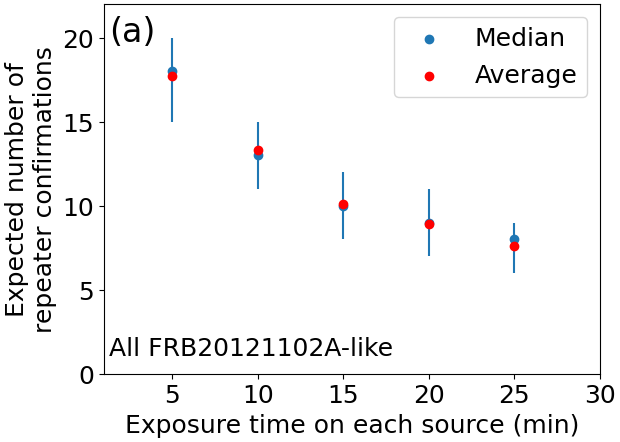}
        \includegraphics[width=0.45\textwidth]{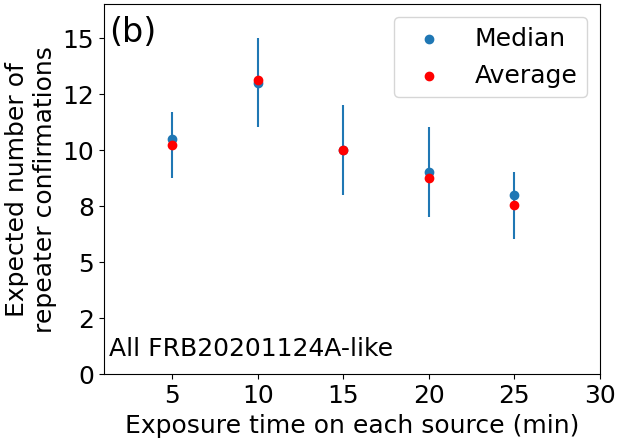}
    \caption{
    The expected number of repeater confirmations as a function of exposure time on each source derived by Monte Carlo simulations (described in Section \ref{sec:exptime}) in twelve hours of observation time.
    (Left) The observed parameters of actively repeating FRB~20121102A are assumed.
    Each blue (red) dot indicates the median (average) value of the number of repeater confirmations over the 10,000 times Monte Carlo simulations for a given exposure time on each source.
    The blue vertical error bars correspond to $\pm 1 \sigma$ data distributions of the Monte Carlo simulations.
    (Right) Same as left, but assuming FRB~20201124A-like repeating parameters. 
    The duty cycle of FRB 20121102A is assumed.
    }
    \label{fig3}
\end{figure*}

Fig. \ref{fig3} (left) shows the anticipated number of repeater confirmations as a function of the exposure time allocated to each source. 
These Monte Carlo simulations are based on the observed parameters of FRB~20121102A.
The results indicate that shorter exposure times per source increase the likelihood of confirming more repeating FRB sources.
This is because even a brief exposure time (e.g., 5-10 minutes) can yield more than one FRB detection if the observatoin coincides with the source's active phase, assuming a repeating rate of 28 FRBs hr$^{-1}$.
Identifying a repeater source requires only a single FRB detection. Therefore, allocating shorter exposure times across a greater number of FRB sources increases the likelihood of confirming repeaters, compared to dedicating longer exposures to fewer sources.

FRB~20201124A represents another active repeater observed with FAST \citep[e.g.][]{ATel14518}. 
\citet{ATel14518} reports over 100 FRB detections from MJD 59305 to 59310, implying an active duration exceeding 6 days. 
Using the total exposure time of 14 hours at the source\footnote[2]{\url{https://fast.bao.ac.cn/observation_log/observed_source_search}}, we infer a repeating rate of 7 hr$^{-1}$ during its active phase. 
As there have been no specific report on the duty cycles of FRB~20201124A to date, we use the duty cycle of FRB~20121102A \citep[i.e., 63.6\%;][]{ATel13959} as reference. This assumption suggests a pattern where FRB 20201124 remains inactive for approximately 9.4 days followed by a 6-day active phase. The results of the Monte Carlo simulations, using parameters analogous to FRB~20201124A, are shown in Fig. \ref{fig3} (right). 
An exposure time of 10 minutes per source yields the highest number of repeater confirmations.

Monte Carlo simulations based on FRB~20121102A and FRB~20201124A indicate that a 10-minute exposure time per source optimizes the likelihood of confirming repeaters. Therefore, we adopted a 10-minute exposure time for each source in our FAST observations (Section \ref{sec:observation}). The simulations estimate that approximately ten repeating FRB sources could be confirmed, assuming that all sources observed in this study are active repeaters similar to FRB~20121102A or FRB~20201124A. 

\begin{figure*}
	\includegraphics[width=\textwidth]{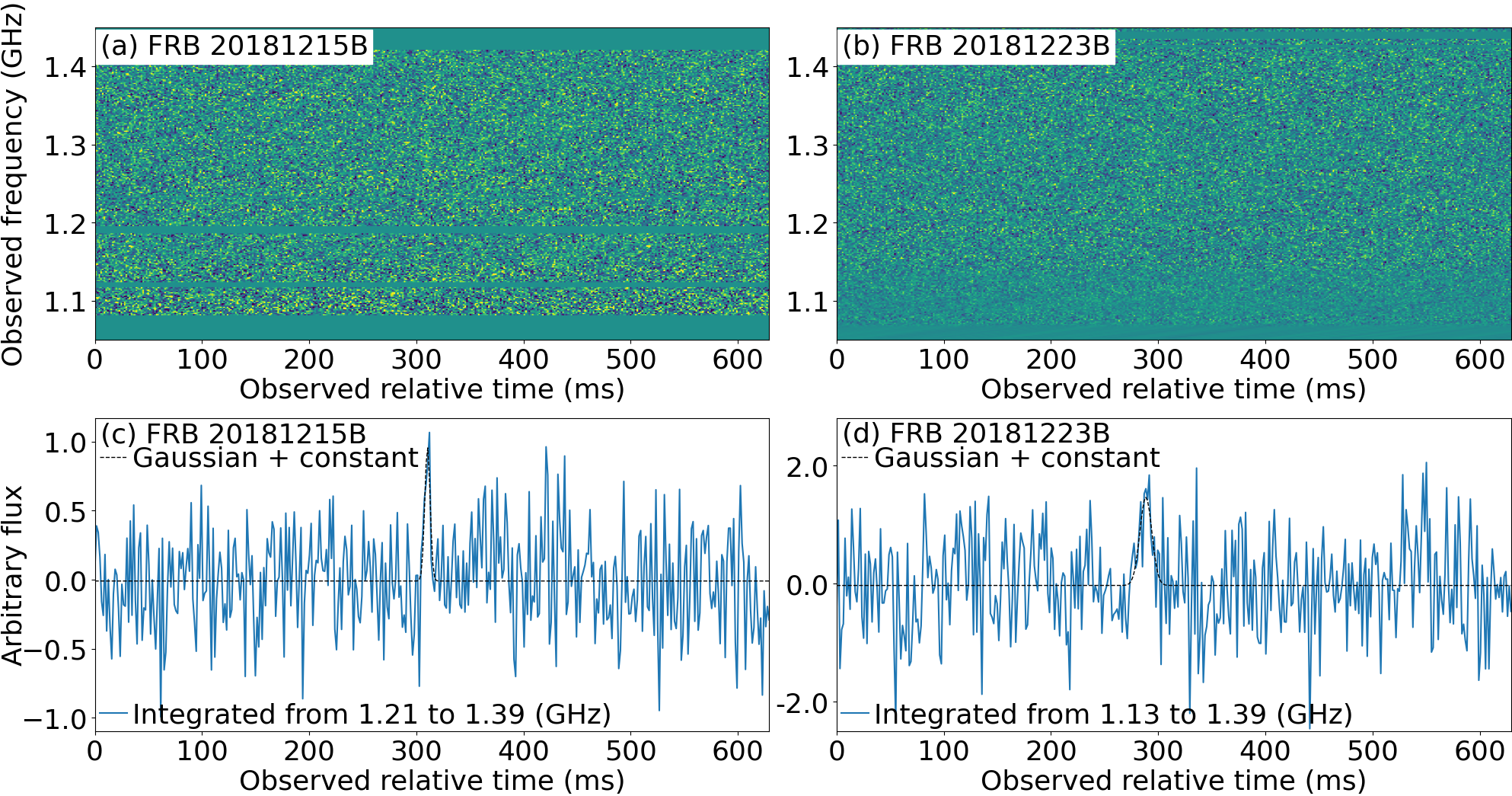}
    \caption{
    (Top) Waterfall plots of the visually selected pulse candidates from the data of (a) FRB~20181215B and (b) FRB~20181223B. 
    The dynamic spectra were dedispersed using DM values of 494.0142 pc cm$^{-3}$ for FRB~20181215B and 565.655 pc cm$^{-3}$ for FRB~20181223B, as reported in \citet{CHIME2021_first}.
    (Bottom) Light curves (solid blue lines) obtained by integrating the waterfall plots for (c) FRB~20181215B over the frequency range 1.21--1.39~GHz and for (d) FRB~20181223B over 1.13--1.39~GHz.
    The dashed black line in each panel represent the best fit of a single Gaussian function with a constant background parameter to the measured light curve.
    The derived S/N values for the candidate signals are 3.9 and 4.0 for FRB~20181215B and FRB~20181223B, respectively.
    }
    \label{fig4}
\end{figure*}

\section{Observation and Analysis}
\label{sec:observation}
We conducted follow-up observations of 36 repeater candidates, selected based on the criteria outlined in Section \ref{sec:target_selection}, using FAST. 
A summary of the observed sources is provided in Tab. \ref{tab:frb-observations} while the telescope specifications are detailed in Tab. \ref{tab:key-properties}.
The observations were centred at 1.25 GHz in the L~band, covering 1.05-1.45~GHz, and employed the pulsar backend in tracking mode. 
The configurations included full polarisations, a sampling time of 196.608 $\mu$s, and 8192 channels. 
Each observational block targeted a single source with a 10-minute exposure, scheduled between 4 September 2021 and 22 June 2022, depending on source visibility.

We searched for bursts up to 200 ms in width. The number of channels used for the incoherent dedispersion is 64. Each bin includes 128 frquency channels. 
Our targets have a median smearing time of $\sim$0.12 ms.
We calculated the sensitivity losses of our targets using the individual burst duration and DMs, following \citet{2015MNRAS.452.1254K}. The typical sensitivity loss is about 0.1\%.
The number of polarisations used in our analysis is $n_{p}=2$.
After the elimination of radio frequency interference 
by deploying \texttt{rfifind} on PRESTO\citep{2011ascl.soft07017R}, pulse candidates were identified with a S/N threshold of 7. 
Since the DMs of all targets are known, a lower threshold was regarded as 
adequate to identify FRBs compared to the typical S/N of 10.

Waterfall plots were generated for each target, focusing on 2.5-second intervals around single-pulse candidates, facilitating thorough visual inspection.
In cases where the number of single-pulse candidates for a target exceeded 1,000, we generated a comprehensive waterfall plot covering the full 10-minute exposure time and conducted a detailed visual inspection. However, neither approach revealed any significant signals.

To convey the depth of our visual inspections, we present the two most significant candidates from FRB20181215B and FRB~20181223B in Fig. \ref{fig4},
selected from our data. The dynamic spectra for FRB~20181215B and FRB~20181223B were dedispersed using DM values of 494.0142 pc cm$^{-3}$ and 565.655 pc cm$^{-3}$, respectively \citep{CHIME2021_first}, to generate the corresponding waterfall plots (Fig. \ref{fig4} top). The waterfall plots were integrated over frequency ranges of 1.21--1.39 GHz for FRB~20181215B and 1.13--1.39 GHz for FRB~20181223B to produce the light curves (Fig. \ref{fig4}, bottom).

We fit a single Gaussian function with a constant background parameter (dashed black lines in Fig. \ref{fig4}, bottom) to the peak of each light curve to derive the S/N.
The S/N values are estimated to be 3.9 and 4.0 for candidates in the FRB~20181215B and FRB~20181223B data, respectively. 
Based on these results, we conclude that there is no significant detection in our data with S/N $>$ 7.

\begin{table*}
\caption{
A summary of burst properties of the 36 targets. Rerr represents the error in Right Ascension, measured in degrees, while Derr indicates the error in Declination, also measured in degrees. Note that this Rerr is converted to $\delta$ (hour angle) $\times \cos{\rm (Decl.)}$ so that it can be compared with Derr in degree (Decl.) with the same scale. Rerr for FRB~20181214F is sourced from CHIME/FRB Catalogue Version 1, while errors for other coordinates from CHIME are taken from CHIME/FRB Catalogue Version 2.
 $W_i$ is intrinsic burst width. 
 The $W_i$ of CHIME sources represents the {\fontfamily{qcr}\selectfont
width\_fitb}, with errors calculated as the sum of {\fontfamily{qcr}\selectfont
width\_fitb\_err} for these bursts. The $W_i$ of FRB~20130729 is the reported intrinsic width of $<$ 4 ms. The $W_i$ values for other FRB sources were derived using Eq. (\ref{eq:wb}), based on observed widths reported in the literature, assuming zero scattering where specified. For FRB~20110523A, the unscattered pulse width was 1.73~ms \citep{2015Natur.528..523M}. The observed widths for FRB~20110703A ($<$ 4.3 ms), FRB~20160410A (4 ms), FRB~20171213A (1.5 ms), FRB~20180212A (1.81 ms) were used to derive their intrinsic widths, assuming no scattering \citep{2013Sci...341...53T, 2017MNRAS.468.3746C, 2018Natur.562..386S}. For FRB~20151230A, the observed width, with scattering effects removed, was reported as 4.4 $\pm$ 0.5 ms \citep{2018MNRAS.475.1427B}. Finally, the intrinsic width of FRB~20010621A, also known as J1852$-$08, was derived from an observed width of 7~ms.
$t_{\rm scatt \, o}$ is the scattering time reported by original papers. 
$t_{\rm scatt \, o}$ of FRB~20151230A is scaled to 1382~MHz, the center frequency of P(b), from the reported scattering time of 18$\pm6$~ms at 1~GHz \citep{2018MNRAS.475.1427B}. 
Similarly, the $t_{\rm scatt \, o}$ of FRB~20130729A is scaled to 1352 MHz from the reported scattering time of 23$\pm2$~ms at 1~GHz \citep{2016MNRAS.460L..30C, 2011MNRAS.415.3065K}.
 $t_{\rm scatt\,FAST}$ is the scattering time scaled to 1.25 GHz.
$T_{P}$ is exposure time in the original observations in hours. Tel denotes the telescopes that were used for observation, where `C' represents CHIME, `G' stands for GBT, `P' corresponds to Parkes, 
and letters following `P' represent different observational datasets, with the corresponding references as follows: (a) \citet{2013Sci...341...53T}, (b) \citet{2018MNRAS.475.1427B}, (c) \citet{2016MNRAS.460L..30C}, and (d) \citet{2011MNRAS.415.3065K}.  `U' signifies UTMOST, and `A' is for ASKAP.
Sel is an abbreviation for the term `Selection.' In the Selection column, `E' denotes empirical selection methods \lq FRBCAT\_empirical\rq and
\lq CHIME\_empirical\rq, while `M' represents machine learning selection methods \lq CHIME\_ML\rq as defined in Section \ref{sec:target_selection}. Properties that were not reported in the original FRB discovery papers are left blank.
}
\begin{tabular}{|l|l|l|l|l|l|l|l|l|l|l|l|}
\label{tab:frb-observations} \\
\hline 
FRB ID  & RA & Rerr & Dec  & Derr  &DM  &  $W_i$ & $t_{\rm scatt\,o}$  & $t_{\rm scatt\,FAST}$  &  $T_{P}$ & Tel &Sel \\
 & (J2000) &  (deg)  &  (J2000) & (deg) &(pc cm$^{-3}$) &  (ms) &  (ms) & (ms) & (hr) &   &     \\
\hline 

20190129A & 03:00:24.00 & 0.196
& $+$21:24:00.0 & 0.23 & 484.76 (1) & 1.13 (30)& 10 (2) & 0.5 (1) &  13.4 (73) & C& M \\
20190131B & 23:38:48.00 &0.235
& $+$11:42:00.0 &0.23& 1805.729 (4) & 0.920 (76)& $<$1.1 & $<$0.06 & 13.5 (58) & C& M \\
20190221A & 08:50:24.00 & 0.049
& $+$09:54:00.0 & 0.26&223.81 (2) & 0.97 (13)& 0.41 (8) & 0.022 (4)& 12.0 (68) & C & E\\
20190226C & 01:10:00.00 & 0.188
&$+$26:45:36.0 & 0.056 &827.77 (2) & 1.31 (12) & $<$ 1.5 & $<$ 0.08 & 4.1 (58) & C& M \\
20190601B & 01:11:36.00 & 0.156
& $+$23:49:12.0 &0.036& 787.80 (9) & 4.04 (36) & 5.7 (3) & 0.30 (1) & 9.6 (73) & C& M \\
20190604G & 08:03:14.40 & 0.012
& $+$59:30:00.0 &0.15& 233.05 (1) & 1.190 (98) & 0.31 (5) & 0.017 (3) & 27 (12) & C& E\\
20190605D & 01:46:48.00 & 0.202
& $+$28:36:00.0 &0.25& 1656.533 (6) & 1.069 (87) & $<$1.2 & $<$0.06 &  11.4 (69) & C& M \\
20110523A & 21:45:12.00 &--- & $-$00:09:37.0 &---& 623.30 (6) & 1.304 &1.66 (14)  &  0.28 (2) & 0.292 & G& E\\
20110703A & 23:30:51.00 & --- & $-$02:52:24.0 & --- & 1103.6 (7) & < 4.080 & --- & ---& 0.075 & P (a) & E\\
20160410A & 08:41:25.00 & ---  & $+$06:05:05.0 &---& 278 (3) & 3.928 & --- & --- & 0.5 &U& E \\
20171213A & 03:39:00.00 & 0.491 

& $-$10:56:00.0 &0.333& 158.6 (2) & 0.579 & --- & --- & 35.5 & A& E\\ 
20181203A & 02:14:24.00 &  0.025
& $+$23:36:00.0 &0.19& 635.926 (2) & 0.171 (84) & 1.4 (2) & 0.08 (1) & 14.9 (66) & C& M\\
20181231A & 01:58:48.00 &  0.159
& $+$21:00:00.0 &0.19& 1376.732 (2) & 0.705 (25) & $<$0.76 & $<$0.04  & 15.0 (64) & C& M\\
20190102A & 00:37:12.00 &0.161
& $+$26:43:12.0 &0.057& 699.173 (6) & 0.824 (73) & 0.99 (7) & 0.052 (4) & 11.5 (81) & C& M\\
20190118B & 02:38:48.00 & 0.229
& $+$23:36:00.0 &0.27& 670.9 (2) & 3.7 (17) & 19 (5) & 1.0 (2) & 12.4 (78) & C&M\\
20151230A & 09:40:50.00 & ---  & $-$03:27:05.0 &---& 960.4 (5) & 4.238 & 4.9 (16) & 7.4 (25) & 0.15   & P (b)& E\\
20190428A & 11:22:55.20 &0.025
& $+$23:18:00.0 &0.15& 969.400 (2) & 0.374 (66) & 3.6 (4) & 0.19 (2) & 16.1 (47) & C& M\\
20181223B & 11:39:36.00 & 0.205
& $+$21:36:00.0 &0.24& 565.66 (2) & 1.57 (31) & 4 (1) & 0.19 (6) & 12.4 (72) & C& M\\
20190228A & 12:13:55.20 &0.005
& $+$22:54:00.0 &0.12 & 419.083 (5) & 2.25 (10) & 18.9 (8) & 1.00 (4) & 16.0 (50)  & C& M \\
20130729A & 13:41:21.00 &---& $-$05:59:43.0 &---& 861 (2) & < 4 & 6.9 (6) & 9.4 (8)&  0.075  & P (c) & E\\
20180212A & 14:21:00.00 &0.499 
& $-$03:35:00.0 & 0.5 &  167.5 (5) & 1.150 & --- & --- & 525.6 & A & E\\
20181128D & 14:22:28.80 &0.016
& $+$59:54:00.0 &0.12 & 146.501 (4) & 1.071 (42) & 0.106 (1) & 0.00562 (7) &  30.9 (67) & C& E\\
20190111A & 14:28:00.00&0.134
& $+$26:48:00.0 & 0.12& 171.9682 (8) & < 1.588 (45) & 0.54 (3) & 0.028 (1) & 16.2 (51)  &C& M\\ 
20181022E & 14:44:48.00 &0.187
& $+$27:06:00.0 &  0.22&285.99 (1) & 0.40 (15) & 0.63 (9) & 0.035 (5) & 13.3 (77) & C& M \\
20190304C & 14:52:00.00 &0.205
& $+$26:42:00.0 & 0.25& 564.99 (1) & 0.948 (89) & $<$1.1 & $<$0.06 & 12.5 (79) & C& M\\
20180915B & 15:00:48.00 &0.227
& $+$25:00:00.0 &0.23 & 177.13 (2) & 1.694 (75) & 0.111 (8) & 0.0059 (4) & 13.1 (76) &C& E, M \\
20181221A & 15:22:24.00 & 0.180
& $+$25:54:00.0 & 0.21& 316.24 (1) & 0.754 (83) & 1.32 (7) & 0.070 (4) & 13.4 (76) & C& M \\
20190125B & 15:26:00.00& 0.159
& $+$50:30:00.0 &0.23 & 178.24 (2) & 2.47 (16) & 1.0 (3) & 0.05 (1) & 17 (11) & C& E, M\\
20181127A & 16:15:12.00 & 0.208
& $+$25:24:00.0 &0.25& 930.32 (1) & 0.74 (13) & 0.58 (9) & 0.031 (5) & 11.8 (80) & C& M \\
20190110C & 16:28:00.00 & 0.142
& $+$41:25:12.0 &0.050& 221.96 (2) & 0.390 (76) & 0.22 (3) & 0.012 (2) & 12.4 (97) & C& M \\
20181228B & 16:41:36.00 &0.092
& $+$63:54:00.0 &0.21& 568.651 (2) & $<$0.1 & 1.16 (8) & 0.062 (4) &  27 (15) & C& M \\
20181214F & 16:50:28.80 & 0.040
& $+$32:24:00.0 &0.22& 2105.8 (1) & 2.3 (56) & 1.5 (3) & 0.08 (2) & 11.4 (67) & C& M \\
20190118A & 16:53:14.40 & 0.028
& $+$11:33:00.0 &0.084 & 225.1080 (7) & 0.140 (16) & 0.282 (6) & 0.0150 (3) & 11.6 (69) & C& M \\
20181215B & 16:59:12.00 & 0.115
& $+$47:33:36.0 &0.030& 494.014 (2) & 0.559 (27) & $<$0.61 & $<$0.03 & 14.1 (9.5) & C& M \\
20190425A & 17:02:48.00 & 0.130
& $+$21:30:00.0 &0.18& 128.1577 (2) & 0.3799 (20) & $<$0.38 & $<$0.02 & 16.9 (42) & C&M \\
20010621A & 18:52:05.00 &---& $-$08:29:35.0 &---& 745 (10) & 5.853 & <3 & <4.38 
& 4.2 & P (d)& E \\
\hline
\end{tabular}
\end{table*}

\begin{table*}
    \caption{Key properties of FAST, CHIME, Arecibo, GBT, Parkes, UTMOST, and ASKAP observations. The channel width is determined by dividing the bandwidth by the number of channels. 
     The terms $T_{\rm sys}$, $G$, $n_p$, and $\Delta \nu$ in Eq.  (\ref{eq:sensitivity}) correspond to the system temperature, antenna gain, number of polarisations, and frequency bandwidth, respectively. 
     In Eq. (\ref{eq:wb}), the term $t_{\rm samp}$  represents the sampling time. In Eq. (\ref{eq:tchan}), $\Delta \nu_{\rm chan}$ and $\nu$ denote the channel width and central observed frequency, respectively.
     The specifications for each telescope were sourced from the following references: 
     CHIME from \citet{CHIME2021_first}, 
     Arecibo from \citet{Good2023, 2013ApJ...775...51D},
     GBT from \citet{2015Natur.528..523M}, 
     Parkes (a) from \citet{2013Sci...341...53T} 
     Parkes (b) from \citet{2020MNRAS.493.1165M,2018MNRAS.475.1427B,2010MNRAS.409..619K} 
     Parkes (c) from \citet{2010MNRAS.409..619K, 2016MNRAS.460L..30C} 
     Parkes (d) from \citet{2001MNRAS.328...17M, 2011MNRAS.415.3065K, 2010MNRAS.401.1057K},
     UTMOST from \citet{2017MNRAS.468.3746C}, and
     ASKAP from \citet{2018Natur.562..386S, 2017ApJ...841L..12B, 2019ApJ...887L..30K}. $^*$~The S/N of the CHIME repeater is 8.5 \citep{2023ApJ...947...83C}.
     }
    
    \begin{tabular}{|l|l|l|l|l|l|l|l|l|l|l|}
    \hline
     & FAST& CHIME & Arecibo & GBT & Parkes (a) & Parkes (b) & Parkes (c) & Parkes (d) & UTMOST & ASKAP \\
    \hline
     Centre Frequency (MHz)& 1250 & 600  & 327 & 800 & 1382 &1382&1352&1374& 843& 1320\\
     Frequency Bandwidth (MHz) &
     400  & 400 & 50& 200 & 400 &400 &340&288&16& 336 \\
     System Temperature (K) & 23 & $\sim$ 50 & 115& 1.16 & 21  &23&23&21&400& 50\\
     Antenna Gain (K/Jy) & 16 &1.16 & 10 & 2 & 0.735&0.735&0.735&0.735 & 3 & 0.1\\
     Sampling Time ($\mu$s) & 196.608& 983 & 80& 1024 &64&64&64&250&655.36&126\\
     Number of polarisation & 2
     & 2 & 2 & 2 & 2 &2&2&2&1&2\\
     Channel Width (MHz) 
     & 0.05 
     & 0.391 
     & 0.025 
     & 0.049 
     & 0.391 
     & 0.391 
     & 0.391
     & 1    
     & 0.098 
     & 1 \\ 
     S/N threshold & 7 & $12^*$ & 6& 8 & 9 &8 &10&8& 10&9.5\\
    \hline 
    \end{tabular}
    \label{tab:key-properties} 
\end{table*}

\renewcommand{\arraystretch}{1.5}
\begin{table*}
\caption{
Overview of sensitivity, fluence, positional accuracy, and repeating rate upper limits.
$S_P$ is sensitivity limit of prior observations at the time of burst detection, while $S_F$ is the 7$\sigma$ sensitivity limit of our FAST observations. 
FL$_{\rm min}$ is the 7$\sigma$ fluence limit in Jy~ms.
$P_{\rm acc}$ is positional accuracy, as detailed in subsection \ref{sub:upperlimit}. 
$r_{\rm scaled}$ denotes the upper limit on the scaled repeating rate, estimated using two statistical models. $r_{\text{scaled}}(P)$ is based on a Poisson distribution, while $r_{\text{scaled}}(W)$ uses a Weibull distribution. These are indicated by the grey, blue, and red triangles in Figs. \ref{fig5} and \ref{fig6}.
These are indicated by the blue and red triangles in Figs. \ref{fig5} and \ref{fig6}. 
The $k$ defines the shape of the Weibull distribution. The $r_{\rm scaled}$ and $k$ values
of FRB~20190111A in "This work" is the combined upper limits from CHIME, Arecibo and FAST observations. Similarly, for FRB~20190110C in "This work," they correspond to the combined upper limits from CHIME, CHIME follow-up and FAST observations. For other sources, these values represent the combined rate from prior and FAST observations. The uncertainties are expressed as 90\% confidence interval. }

\begin{tabular}{|l|l|l|l|l|l|l|l|l|l|l|}
\label{tab:result} \\
\hline 
 & & & & &\multicolumn{3}{c|}{CHIME+}  & \multicolumn{3}{c|}{This work} \\
FRB ID &$S_{\rm P}$ & $S_{\rm F}$&  FL$_{\rm min}$ &$P_{\rm acc}$&$r_{\rm scaled} (P)$ & $r_{\rm scaled} (W)$ &$k$&$r_{\rm scaled} (P)$ & $r_{\rm scaled} (W)$ & $k$\\
 &  (Jy) & (Jy) & (Jy ms) & & (hr$^{-1}$) & (hr$^{-1}$) &  &(hr$^{-1}$)  & (hr$^{-1}$) &\\
\hline 

20190129A 
&1.818
&0.011
&0.014
&0.109
&0.18$^{+0.54}_{-0.17}$
&0.13$^{+0.53}_{-0.12}$
&1.4$^{+6.8}_{-1.2}$
&0.048$^{+ 0.14}_{-0.044}$
&0.045$^{+0.57}_{-0.042}$
&0.56$^{+6.3}_{-0.41}$
\\

20190131B 
&3.276
&0.012
&0.013
&0.096
&0.44$^{+ 1.29}_{-0.40}$
&0.32$^{+1.2}_{-0.30}$
&1.4$^{+6.8}_{-1.2}$
&0.071$^{+ 0.21}_{-0.066}$ 
&0.088$^{+1.5}_{-0.083}$
&0.45$^{+5.8}_{-0.3}$
\\

20190221A 
&1.140
&0.012
&0.011
&0.112
&0.10$^{+0.30}_{-0.09}$
&0.074$^{0.29}_{0.069}$
&1.4$^{+6.8}_{-1.2}$
&0.040$^{+ 0.12}_{-0.037}$
&0.032$^{+0.28}_{-0.030}$
&0.72$^{+6.6}_{-0.56}$

\\

20190226C 
&1.568
&0.010
&0.013
&0.115
&0.48$^{+1.4}_{-0.44}$
&0.35$^{+1.4}_{-0.33}$
&1.4$^{+6.8}_{-1.2}$
&0.047$^{+ 0.14}_{-0.043}$
&0.089$^{+1.7}_{-0.085}$
&0.37$^{+5.1}_{-0.23}$
\\

20190601B 
&0.531
&0.006
&0.023
&0.137
&0.040$^{+0.12}_{-0.037}$
&0.030$^{+0.12}_{-0.027}$
&1.4$^{+6.8}_{-1.2}$
&0.013$^{+ 0.037}_{-0.012}$
&0.011$^{+0.12}_{-0.01}$
&0.62$^{+6.4}_{-0.46}$
\\

20190604G 
&0.952
&0.010
&0.013
&0.094
&0.034$^{+0.10}_{-0.032}$
&0.025$^{+0.1}_{-0.023}$
&1.4$^{+6.8}_{-1.2}$
&0.023$^{+ 0.066}_{-0.021}$
&0.017$^{+0.082}_{-0.016}$
&1.1$^{+6.8}_{-0.90}$
\\

20190605D 
&2.701
&0.011
&0.013
&0.092
&0.39$^{+1.1}_{-0.36}$
&0.28$^{+1.1}_{-0.27}$
&1.4$^{+6.8}_{-1.2}$
&0.065$^{+ 0.19}_{-0.059}$
&0.079$^{+1.3}_{-0.074}$
&0.45$^{+5.8}_{-0.31}$
\\

20110523A 
&0.009 
&0.010
&0.014
&0.179
&0.0028$^{+0.0082}_{-0.0026}$
& ---
&---
&0.0025$^{+ 0.0075}_{-0.0024}$
& ---
&---
\\

20110703A 
&0.146
&0.006
&0.023
&0.200
&0.74$^{+2.2}_{-0.68}$
& ---
&---
&0.012$^{+ 0.036}_{-0.011}$
& ---
&---
\\

20160410A 
&3.840
&0.006
&0.022
&0.021
&15$^{+ 44}_{-14 }$
& ---
&---
&0.12$^{+ 0.36}_{-0.11}$
& ---
&---
\\

20171213A 
&12.264
&0.015
&0.014
&0.055
&$1.2^{+3.5}_{-1.1}$
& ---
&---
&
0.18$^{+ 0.52}_{-0.16}$
& ---
&---

\\

20181203A 
&10.535
&0.036
&0.011
&0.129
&$2.3^{+ 6.7}_{-2.1 }$
&1.6$^{+5.1}_{-1.5}$
&1.5$^{+6.8}_{-1.3}$
&0.28$^{+ 0.83}_{-0.26}$
&0.40$^{+4.8}_{-0.37}$
&0.42$^{+5.5}_{-0.27}$
\\

20181231A 
&3.733
&0.014
&0.011
&0.123
&$0.48^{+1.4}_{-0.44}$
&0.35$^{+1.4}_{-0.33}$
&1.4$^{+6.8}_{-1.2}$
&0.070$^{+ 0.21}_{-0.064}$
&0.094$^{+1.6}_{-0.089}$
&0.43$^{+5.6}_{-0.29}$
\\

20190102A 
&2.287
&0.013
&0.011
&0.090
&$0.30^{+0.88}_{-0.28}$
&0.22$^{+0.86}_{-0.20}$
&1.4$^{+6.8}_{-1.2}$
&0.072$^{+ 0.21}_{-0.066}$
&0.07$^{+0.94}_{-0.066}$
&0.53$^{+6.2}_{-0.38}$
\\

20190118B 
&0.736
&0.006
&0.023
&0.079
&0.051$^{+0.15}_{-0.047}$
&0.037$^{+0.15}_{-0.035}$
&1.4$^{+6.8}_{-1.2}$
&0.021$^{+ 0.061}_{-0.019}$
&0.016$^{+0.14}_{-0.015}$
&0.73$^{+6.6}_{-0.57}$
\\

20151230A 
&0.170
&0.008
&0.066
&0.179
&0.47$^{+ 1.4}_{- 0.43}$
& ---
&---
& 0.022$^{+ 0.064}_{-0.020}$
& ---
&---
\\

20190428A 
&6.000
&0.021
&0.011
&0.131
&0.91$^{+ 2.7}_{- 0.84}$
&0.66$^{+2.5}_{-0.62}$
&1.4$^{+6.8}_{-1.2}$
&0.13$^{+ 0.37}_{-0.11}$
&0.17$^{+2.7}_{-0.16}$
&0.42$^{+5.6}_{-0.28}$
\\

20181223B 
&1.127
&0.009
&0.014
&0.101
&0.097$^{+0.28}_{-0.088}$
&0.071$^{+0.28}_{-0.066}$
&1.4$^{+6.8}_{-1.2}$
&0.034$^{+ 0.098}_{-0.031}$
&0.028$^{+0.28}_{-0.026}$
&0.66$^{+6.5}_{-0.5}$
\\

20190228A 
&1.152
&0.008
&0.019
&0.129
&0.077$^{+0.23}_{-0.071}$
&0.057$^{+0.22}_{-0.053}$
&1.4$^{+6.8}_{-1.2}$
&0.023$^{+ 0.067}_{-0.021}$
&0.02$^{+0.24}_{-0.019}$
&0.60$^{+6.4}_{-0.44}$
\\

20130729A 
&0.270
&0.009
&0.092 
&0.179 
& 1.9$^{+ 5.5}_{- 1.7}$
& ---
&---
& 0.028$^{+ 0.083}_{-0.026}$
& ---
&---
\\

20180212A 
&6.777
&0.011
&0.012
&0.036
&0.034$^{+0.098}_{-0.031}$
& ---
&---
& 0.028$^{+ 0.083}_{-0.026}$
& ---
&---
\\

20181128D 
&0.877
&0.011 
&0.012
&0.100
& 0.027$^{+0.078}_{-0.024}$
&0.02$^{+0.077}_{-0.018}$
&1.4$^{+6.8}_{-1.2}$
& 0.019$^{+ 0.056}_{-0.018}$
&0.014$^{+0.063}_{-0.013}$
&1.2$^{+6.9}_{-0.98}$
\\

20190111A 
&0.654
&0.009 
&0.014
&0.145
& 0.033$^{+0.096}_{-0.030}$
&0.024$^{+0.095}_{-0.022}$
&1.4$^{+6.8}_{-1.2}$
&0.017$^{+ 0.050}_{-0.016}$
&0.0050$^{+0.021}_{-0.0050}$
&1.2$^{+6.9}_{-0.10}$
\\

20181022E 
&3.056
&0.019 
&0.009
&0.114
& 0.40$^{+1.2}_{-0.37}$
&0.29$^{+1.1}_{-0.27}$
&1.4$^{+6.8}_{-1.2}$
&0.10$^{+ 0.30}_{-0.093}$
&0.096$^{+1.2}_{-0.090}$
&0.55$^{+6.2}_{-0.40}$
\\

20190304C 
&1.795
&0.012 
&0.011
&0.092
&0.19$^{+0.56}_{- 0.18}$
&0.14$^{+0.55}_{-0.13}$
&1.4$^{+6.8}_{-1.2}$
&0.058$^{+ 0.17}_{-0.053}$
&0.051$^{+0.57}_{-0.047}$
&0.6$^{+6.4}_{-0.45}$
\\

20180915B 
&0.621
&0.009 
&0.015
&0.092
& 0.037$^{+ 0.11}_{-0.034}$
&0.027$^{+0.11}_{-0.025}$
&1.4$^{+6.8}_{-1.2}$
&0.022$^{+ 0.064}_{-0.020}$
&0.016$^{+0.092}_{-0.015}$
&0.97$^{+6.8}_{-0.79}$
\\

20181221A 
&1.729
&0.013 
&0.010
&0.114
& 0.17$^{+ 0.50}_{-0.16}$
&0.12$^{+0.49}_{-0.12}$
&1.4$^{+6.8}_{-1.2}$
& 0.054$^{+ 0.16}_{-0.050}$
&0.047$^{+0.50}_{-0.043}$
&0.63$^{+6.5}_{-0.47}$
\\

20190125B 
&0.462
&0.007 
&0.018
&0.092
& 0.018$^{+0.054}_{-0.017}$
&0.014$^{+0.054}_{-0.013}$
&1.4$^{+6.8}_{-1.2}$
& 0.013$^{+ 0.037}_{-0.012}$
&0.0092$^{+0.044}_{-0.0086}$
&1.1$^{+6.9}_{-0.92}$
\\

20181127A 
&2.927
&0.014
&0.011
&0.092
& 0.42$^{+1.2}_{-0.39}$
&0.31$^{+1.2}_{-0.29}$
&1.4$^{+6.8}_{-1.2}$
&0.082$^{+ 0.24}_{-0.075}$
&0.09$^{+1.4}_{-0.085}$
&0.48$^{+6.0}_{-0.33}$
\\

20190110C 
&2.775
&0.019 
&0.008
&0.112
& 0.37$^{+1.1 }_{- 0.34}$
&0.27$^{+1.1}_{-0.25}$
&1.4$^{+6.8}_{-1.2}$
& 0.10$^{+ 0.30}_{-0.09}$
&0.099$^{+0.19}_{-0.069}$
&0.93$^{+6.4}_{-0.66}$
\\

20181228B 
&17.028
&0.057 
&0.015
&0.106
& 2.6$^{+7.6}_{-2.4}$
&1.8$^{+5.4}_{-1.7}$
&1.5$^{+6.8}_{-1.3}$
& 0.60$^{+ 1.8}_{-0.55}$
&0.55$^{+4.8}_{-0.52}$
&0.56$^{+6.2}_{-0.39}$
\\

20181214F 
&1.417
&0.008 
&0.018
&0.127
& 0.15$^{+0.43 }_{-0.14 }$
&0.11$^{+0.43}_{-0.10}$
&1.4$^{+6.8}_{-1.2}$
& 0.025$^{+ 0.074}_{-0.023}$
&0.031$^{+0.52}_{-0.029}$
&0.45$^{+5.8}_{-0.31}$
\\

20190118A 
&7.764
&0.040 
&0.010
&0.123
& 1.9$^{+ 5.5}_{-1.7 }$
&1.3$^{+4.5}_{-1.2}$
&1.5$^{+6.8}_{-1.2}$
& 0.32$^{+ 0.94}_{-0.29}$
&0.36$^{+4.2}_{-0.34}$
&0.47$^{+5.9}_{-0.32}$
\\

20181215B 
&2.838
&0.016
&0.009
&0.165
& 0.34$^{+ 0.99}_{- 0.31}$
&0.25$^{+0.97}_{-0.23}$
&1.4$^{+6.8}_{-1.2}$
& 0.059$^{+ 0.17}_{-0.054}$
&0.07$^{+1.1}_{-0.066}$
&0.46$^{+5.8}_{-0.31}$
\\

20190425A 
&2.271
&0.019 
&0.008
&0.138
& 0.20$^{+ 0.59}_{-0.19 }$
&0.15$^{+0.58}_{-0.14}$
&1.4$^{+6.8}_{-1.2}$
&0.074$^{+ 0.22}_{-0.068}$
&0.061$^{+0.57}_{-0.056}$
&0.68$^{+6.6}_{-0.52}$
\\

20010621A 
&0.136
&0.005
&0.038
&0.201
& 0.012$^{+0.035 }_{- 0.011}$
& ---  
&---
& 
 0.0058$^{+ 0.017}_{-0.0053}$
& ---
& ---
\\

\hline
\end{tabular}
\noindent
\end{table*}
\renewcommand{\arraystretch}{1}  

\section{result}
\label{sec:result}
Our analysis, employing PRESTO and visual inspections, did not reveal any additional bursts from the 36 repeating FRB candidates.
In the following, we estimate the upper limits for the repetition rates of these FRB sources based on our non-detections, building on the analytical framework of \citet{Good2023}.

\subsection{Sensitivity limit}
\label{sensitivity}
We determine the sensitivity limit for each source in our observations, following \citet{ 2003ApJ...596.1142C}, and the equation is as follows:
\begin{equation}
\label{eq:sensitivity}
    S_{\rm min} = \frac{\beta \rm{S/N} (\it{T}_{\rm sys})}{GW_i}\sqrt{\frac{W_b}{n_p \Delta \nu}}.
\end{equation}
Here, $\beta$ represents the digitisation loss.
In this work, we adopt the assumption $\beta=1$ in accordance with \citet{Good2023}. 
$T_{\rm{sys}}$ is systematic temperature in Kelvin, $G$ is antenna gain in units of K~Jy$^{-1}$, $W_i$ is the intrinsic pulse width in seconds, $W_b$ is the broadened pulse width in seconds, $n_p$ is number of polarisations, and $\Delta \nu$ is frequency bandwidth in hertz. 
$W_b$ is given as,
\begin{equation}
\label{eq:wb}
    W_b = \sqrt{W_i^2 + t_{\rm samp}^2 + t_{\rm chan}^2 + t_{\rm scatt}^2},
\end{equation}
where $t_{\rm samp}$ is sampling time (s), and $t_{\rm scatt}$ is scattering time (s).
$t_{\rm chan}$ is smearing described in \citet{2012hpa..book.....L}, and as follows:
\begin{equation}
\label{eq:tchan}
t_{\rm chan} = 8.3 \mu s \bigg( \frac{\Delta \nu_{\rm chan}}{\rm MHz} \bigg) \bigg( \frac{\nu}{\rm GHz} \bigg)^{-3} \bigg( \frac{\rm DM}{\rm pc \, cm^{-3}} \bigg),
\end{equation}
where $\Delta \nu_{\rm chan}$ is frequency channel bandwidth (MHz) and $\nu$ is the central observing frequency (GHz). 

The parameters in Eqs. (\ref{eq:sensitivity}) and (\ref{eq:wb}) for each telescope are summarized in Tabs. \ref{tab:frb-observations} and \ref{tab:key-properties}.
These parameters are used to calculate the sensitivities for both original observations and our FAST observations.
The scattering time is scaled assuming $t_{\rm scatt}\propto \nu^{-4}$ to derive values at 1250~MHz, the frequency at which FAST observations are conducted. 
For some FRBs, no explicit scattering information has been reported. 
In such cases, we assume the scattering value to be zero.

\subsection{Upper limits on the repeating rates}
\label{sub:upperlimit}
We aim to estimate the upper limits on the FRB repeating rates, assuming both the Poisson and Weibull distributions.
The Poisson distribution models random events occurring at a constant average rate, while
the Weibull distribution provides greater flexibility by accommodating clustering events.

\subsubsection{Poisson distribution}
When $N$ events occur randomly over a time period $T$ with an event rate $r$, the probability of observing exactly $N$ events in time $T$ follows a Poisson distribution, 
\begin{equation}
    P(N; rT) = \frac{(r T)^N e^{-r T}}{N!}.
    \label{eq:poisson}
\end{equation}
The event rate in the Poisson process is given by
\begin{equation}
    r = \frac{N}{T}.   
\end{equation} 

We modified the Poisson-scaled repeating rate equation from \citet{Good2023}, which assumes that FRBs occur randomly within a unit of time and includes a sensitivity scaling factor to make them comparable across different sets of observations. Our modification accounts for the probability of source coverage by FAST.
This adaptation allows us to calculate a more acculate upper limit on the repeating rate for each observed source,
($r_{\rm scaled, joint}$):
\begin{equation}
\label{eq:r_scale}
r_{\rm scaled, joint}=\frac{N_{\rm bursts}}{T_{P}(S_{P}/S_{0})^{-1.5}+
P_{\rm acc}
T_{F}(S_{F}/S_{0})^{-1.5}},
\end{equation}
where $N_{\rm bursts}$ is the number of detected bursts, $T_{P}$ is the on-source exposure time in hours, and  $S_{P}$ is the sensitivity limit in Jansky of previous observations (e.g., CHIME and Parkes).
$T_{F}$ and $S_{F}$ are the on-source exposure time and sensitivity of our FAST observations, respectively.
The adopted on-source exposure times and sensitivities are summarised in Tabs. \ref{tab:frb-observations} and \ref{tab:result}, respectively.
$S_{0}$ is the reference flux density of 1~Jy \citep{Good2023}. We adopt the power of $-1.5$ in the flux-density term, which corresponds to the $N_{\rm{burst}} \propto S_{\rm{min}}^{\alpha} (\alpha=-1.5)$ expectation for a non-evolving population in Euclidean space \citep[e.xg.][]{CHIME2021_first}.
The assumed value is consistent with the slope of $\alpha=-1.4\pm0.11$ (statistical error)$^{+0.06}_{-0.09}$ (systematic error) measured by CHIME \citep{CHIME2021_first}.
$P_{\rm acc}$ is the probability 
of the source coverage with FAST, calculated using the following equation, 
\begin{equation}
    \label{eq:positional_accuracy}
    P_{\rm acc} = R_{\rm FAST \, beam}  \times P_{\rm unc},
\end{equation}
where $R_{\rm FAST \, beam}$ denotes the ratio of the overlapped area between FAST's 19-beam configuration (defined by its Full Width at Half Maximum, FWHM) and the positional uncertainty area reported in the original detection papers. 
For a visual representation, see Appendix \ref{sec:appendix_A}, which illustrates the overlap ratio between FAST’s 19-beam configuration and the positional uncertainty region.
For CHIME-discovered FRBs, the positional uncertainty area is represented as a rectangle. 
The width of this rectangle corresponds to the right ascension positional error, denoted as $R_{\rm err}$ in Tab. \ref{tab:frb-observations}, while the height is aligned with the declination positional error, referred to as $D_{\rm err}$ in the Tab. \ref{tab:frb-observations}.
This rectangular representation is chosen because it closely approximates the beam pattern of CHIME, as illustrated in Fig. 6 of \citet{CHIME2021_first}. The positional uncertainty for other FRBs is characterised by elliptical regions. For ASKAP-discovered FRBs, the semi-major and semi-minor axes are represented by $R_{\rm err}$ and $D_{\rm err}$, respectively. For the rest of FRBs, the semi-major and semi-minor axes are derived from the FWHM values reported by the original FRB discovery papers. 

$P_{\rm unc}$ is defined based on the confidence intervals of the uncertainty values reported in the original studies, as the definition of uncertainty varies between works. Specifically, the values assigned to FRBs identified by CHIME, ASKAP, and other sources are 0.68, 0.90, and 0.76, respectively \citep{CHIME2021_first, 2018Natur.562..386S}. The value of 0.76 is derived based on the assumption that the FWHM of the beam pattern follows a Gaussian distribution. 

To estimate confidence intervals for $r_{\rm scaled, joint}$, we employ the Bayesian approach described by \citet{1991ApJ...374..344K}, which is particularly effective for small observed burst counts, offering more realistic upper limits than Gaussian approximations. 

The Poisson upper limits on the repeating rates for the 36 FRB sources, derived using Eq. (\ref{eq:r_scale}), are shown in Fig. \ref{fig5}.

\subsubsection{Weibull distribution}

The activity of FRBs may not strictly follow a Poisson process, and the distribution of intervals between bursts can offer more insight than simply analysing the total number of bursts over time. The Weibull distribution provides a generalised probability density function (PDF) for the distribution of intervals between burst events. 
The PDF of the time interval between events in a Poisson process is given by
\begin{equation}
   p(\delta|r) = re^{-\delta r},
\end{equation}
where $\delta$ represents the time between consecutive events (i.e., the event interval) and $r$ denotes the constant repetition rate. 

The Weibull distribution introduces an additional parameter, $k$, known as the shape parameter, which characterizes the clustering of FRB activity. In this work, we specifically follow the formalism of \citet{2018MNRAS.475.5109O}. The PDF of the Weibull distribution is given by,

\begin{equation}
p(\delta | \tau, k) = \frac{k}{\tau} \left( \frac{\delta}{\tau} \right)^{k-1} e^{-(\frac{\delta}{\tau})^k}
\end{equation}
where $\tau$ is the scale parameter, defined as 
\begin{equation}
\tau = \frac{1}{r\Gamma\left(1 + \frac{1}{k} \right)}
\end{equation}

\noindent with $\Gamma$ denoting the gamma function, given by
\begin{equation}
\Gamma(x) = \int_0^{\infty} dt \, t^{x-1} e^{-t}.
\end{equation}

For $k = 1$, the Weibull distribution reduces to the Poissonian case. When $k \neq 1$, the Weibull distribution deviates from Poisson statistics, introducing a clustering effects. In particular, for $k < 1$, the distribution becomes right-skewed, favouring shorter time intervals between events, leading to bursts occuring in clusters. 

The posterior PDF of the parameters $k$ and $r$, conditioned on the observed data, is given by,
\begin{equation}
    p(k, r \mid N, t_1, \dots, t_N) \propto p(N, t_1, \dots, t_N \mid k, r) p(k, r),
\end{equation}
where $p(k, r)$ is Jeffreys priors given by,
\begin{equation}
    p(k,r) \propto k^{-1}r^{-1}.
\end{equation}

In our calculations, we adopt the approximation from \citet{2018MNRAS.475.5109O}, treating the total likelihood as the product of individual observation likelihoods. This approximation is valid when the time intervals between observations are sufficiently longer than the expected time between successive bursts. 
Unlike \citet{2018MNRAS.475.5109O}, which focused on the analysis of an active repeater, we do not know the typical timescale between successive bursts in our sample. As a result, we cannot strictly verify whether the observation intervals (e.g., $\sim$1 day in the CHIME case) are sufficiently long. Nonetheless, this approximation provides a practical solution for analysing discrete, non-continuous observation campaigns where data are not collected over a single uninterrupted time span.

Under this approximation, it is sufficient to consider only two specific scenarios in our study: the probability of observing either zero or one burst within a fixed continuous observation window. 

The observations were split into individual blocks, and each block had a continuous exposure time on a source.
For instance, CHIME repeats a single-block daily observation, where most blocks include non-detection, and one FRB is detected in one block.  
We then calculate the combined probability by multiplying the individual probabilities. The probability of observing no events during a single finite observational period is given by, 
\begin{equation}
    \label{eq:r_scale_weibull0}
    P(N = 0 | k, r) = \frac{\Gamma_i \left( \frac{1}{k}, (\Delta\, r \, \Gamma(1 + 1/k))^k \right)}{k \, \Gamma(1 + 1/k)},
\end{equation}
where $\Delta$ is the duration of the observational period, expressed in hours throughout this analysis. 
$\Gamma_i$ is the incomplete gamma function, defined as,
\begin{equation}
    \Gamma_i(x, z) = \int_{z}^{\infty} dt\,  t^{x-1} e^{-t} .
\end{equation}

For CHIME~FRB sources, the values of $\Delta$ were calculated for each FRB by dividing the reported total exposure time by 80\% of the 308-day period from 2018 August 28 to 2019 July 1. \citet{Good2023} noted that approximately 80 \% of the observation dates (246 days) were considered suitable for scientific observation, taking into account potential interruptions such as system maintenance and weather. Moreover, since the observation time per day was relatively consistent, we assume that dividing the total exposure time by 80\% of the total observation days provides a reasonable approximation of the actual observation conditions.

Following \citet{Good2023}, $\Delta$ is further scaled by sensitivity in the same manner as in the calculation of the Poisson repetition rate (Eq. \ref{eq:poisson}), 
with an additional scaling factor of $P_{\rm acc}$, i.e.
\begin{equation}
    \Delta_{\rm scaled} = \Delta (S/S_0) ^{-1.5}  P_{\rm acc},
\end{equation}
where $S$ is the sensitivity limit. The factor $P_{\rm acc}$ is applied for FAST observations; otherwise, $P_{\rm acc}$ is set to unity.  
For example, in the case of FRB~20190111A as observed with FAST, $\Delta_{\rm scaled}$ is calculated as follows,
\begin{equation}
    \Delta_{\rm scaled} = 10\,{\rm min} \times \left(\frac{{\rm hour}}{60 \,{\rm min}}\right) \times \left(\frac{0.009}{1}\right)^{-1.5} \times 0.145 = 28.3.
\end{equation}

FRB~20190111A is distinctive in that it has been observed by three telescopes: CHIME, Arecibo, and FAST, necessitating a specialized treatment in our analysis.

\citet{Good2023} reported a CHIME exposure time that exceeds the duration recorded in the catalog, spanning from 2018 August 28 to 2021 May 1 (a total of 978 days). Consequently, we adopt their reported exposure time as the CHIME exposure time for this target. To account for the actual observation dates—approximately 80\% of the total operational days—we consider 782 days of effective observation. Based on this, \citet{Good2023} reported a total CHIME exposure time of 57.5 hours, yielding a $\Delta$ of 57.5/782 hours per day.

For Arecibo observations of the same target, the total exposure time was 3.07 hours over the period from 2019 September 28 to 2019 December 15 \citep{Good2023}. While \citet{Good2023} does not explicitly state the exact number of observation days, private communication with Dr. Good indicates that the observations were likely conducted over approximately five days within this period. Accordingly, we estimate $\Delta$ for Arecibo as 3.07/5 hours per day.

FRB~20190110C is a repeater confirmed at the time of preparing this manuscript \citep{2023ApJ...947...83C}. The follow-up observations had a total exposure time pf 45$\pm11$~hours over a period spanning 2019 September 30 to 2021 May 1 (580~days). Similar to FRB~20190111A, we assume that 80\% of the 580~days were actual observation days, resulting in $\Delta=45/464$ hours per day. During these follow-up observations, 2 additional bursts were detected.

The PDF of observing exactly one burst is given by,
\begin{equation}
    \label{eq:r_scale_weibull1}
    p(N = 1, t_1 | k, r) = r \, \text{CDF}(t_1 | k, r)  \text{CDF}(\Delta - t_1 | k, r),
\end{equation}
where $t_1$ is the time of the first observed burst after the start of a single observation period ($\Delta$). Since the exact time of a burst within an observation session cannot be precisely determined, we set $t_1$ to be half of $\Delta$. Consequently, the burst is assumed to occur at the midpoint of the observation.

The cumulative distribution function (CDF) is defined as
\begin{equation}
    \text{CDF}(\delta \mid k, r) = \exp \left( - [\delta r \Gamma(1 + 1/k)]^k \right).
\end{equation}
The upper limits on the repetition rates for the 28 FRB sources detected by CHIME, assuming a Weibull distribution and derived using Eqs. (\ref{eq:r_scale_weibull0}) and (\ref{eq:r_scale_weibull1}), are shown in Fig. \ref{fig6}.


To obtain the PDF of $\log{r}$, we marginalized the joint PDF over $\log{k}$. The integration ranges used are $10^{-2} < k < 10^{1}$ and $10^{-5} < r < 10^{1}$. 
Within this range, the 50th percentile of the resulting $r$ distribution is reported as the upper limit, with the 5th and 95th percentiles representing the associated uncertainties. The values for $\log{k}$ are obtained similarly, by marginalizing over $\log{r}$.
Examples of the joint PDFs are presented in Appendix \ref{sec:apppendix_b}.

This work follows the formalism established in \citet{2018MNRAS.475.5109O} and \citet{Good2023} with a correction for a minor erratum in the calculation code used in \citet{Good2023}, as clarified through private communications.

Figs. \ref{fig5} and \ref{fig6} represent the calculated upper limits on the repeating rates for the 36 FRB sources, assuming Poisson and Weibull distributions, respectively. The upper limits derived from our FAST observations and prior observations are shown as blue and red triangles, while the results using only prior observations are represented by grey 
triangles. The CHIME-confirmed repeater is 
shown by a magenta circle. For the Poisson distribution, the combined upper limits from FAST observations and prior observations
range from approximately 
10$^{-2.6}$ to 10$^{-0.22}$ hr$^{-1}$. 
For the Weibull distribution, these limits range from about 10$^{-2.3}$ to 10$^{0.25}$ hr$^{-1}$.

\section{discussion}
\label{sec:discussion}
\subsection{Constraints on FRB Repetition Rates and the Weibull Shape Parameter}


\citet{Good2023} utilised Arecibo to conduct follow-up observations for seven apparently non-repeating FRB sources to find no additional FRB.
Their results place upper limits of $r_{\rm scaled} \sim 10^{-1.5}$ hr$^{-1}$ on the repeating rates of the seven samples.
The typical upper limit placed by additional FAST follow-up observations in this work is
$r_{\rm scaled} 
\sim 10^{-1.3}$ hr $^{-1}$ in both Poisson and Weibull cases.
This upper limit is comparable to Arecibo's results \citep{Good2023}, while our upper limits include about five times more samples (i.e., non-repeating FRBs) than those reported by \citet{Good2023}.

In addition to constraining the repetition rate $r$, the Weibull distribution also provides implicit constrants on the shape parameter $k$,  which describes the temporal distribution of burst arrival times. We note that the marginalized posteriors typically have broad 90\% confidence intervals extending up to  $k\sim6$, indicating that $k$ remains poorly constrained in the absence of detected bursts. Neverthless, for many sources, the central values (e.g., 50th percentiles) fall below $k=1$, which may suggest a tendency toward temporally clustered burst behaviour---consistent with previous findings for known repeaterss such as FRB~20121102 \citet{2018MNRAS.475.5109O}. However, both clustered and periodic behaviours remain viable under current data. These results underscore the need for extended and continuous observations, along with improved modelling frameworks that incorporate the discrete sampling inherent in follow-up campaigns.

\subsection{FRB~20190110C}
FRB~20190110C, one of the targets analysed in this study, was originally classified as a non-repeating FRB in the first CHIME~FRB catalogue \citep{CHIME2021_first}. However, in a recent publication, CHIME team reported the discovery of 25 new repeating FRBs \citep{2023ApJ...947...83C}, six of which were also listed in the first catalogue. 
Notably, FRB~20190110C, which include three bursts detected with CHIME, is among these newly identified repeaters. This aligns with its classification by the machine-learning method employed in this work, further validating the approach introduced by \citet{Chen2022}.

\begin{figure*}
    \includegraphics[width=0.9\textwidth]{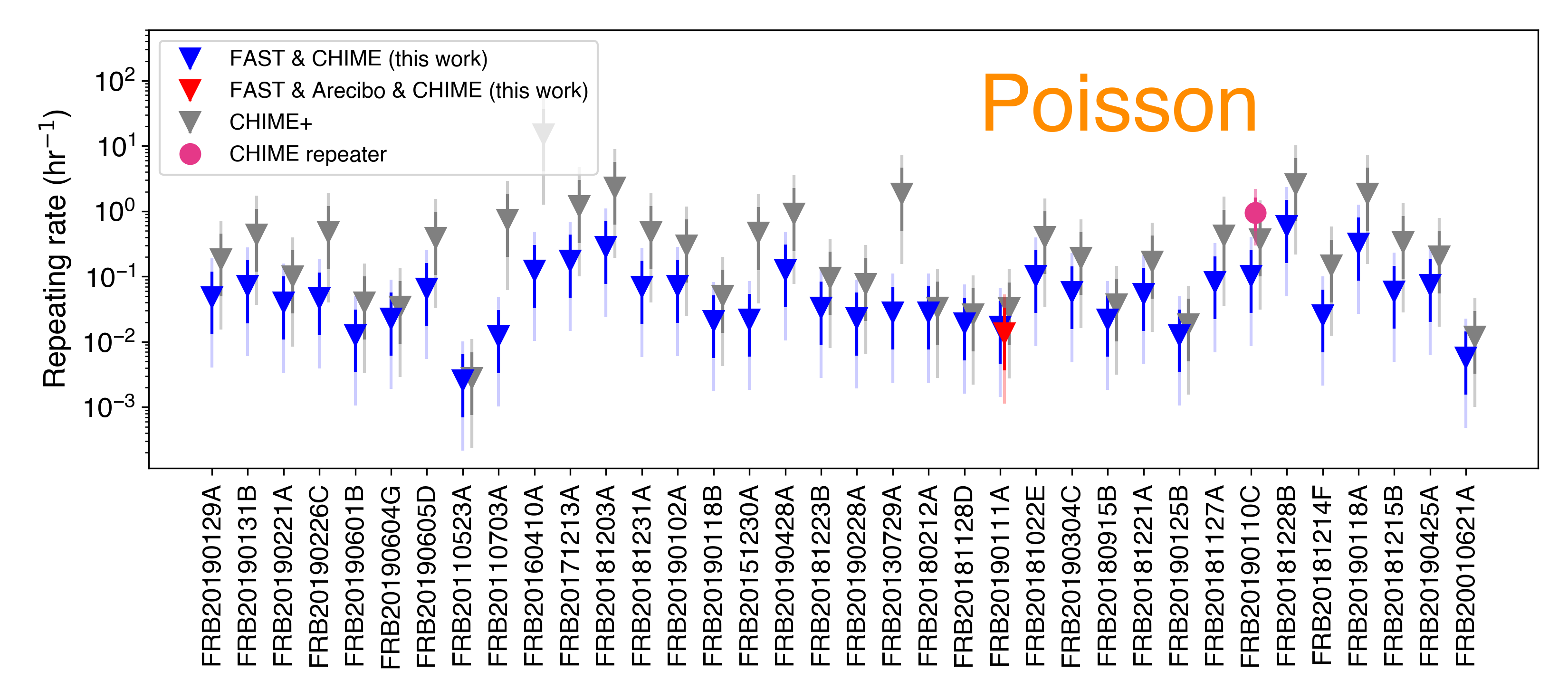}
    \caption{
    Upper limits on Poissonian repeating rate ($r_{\rm scaled, joint}$) of 36 FRB sources derived by Eq. (\ref{eq:r_scale}).
    Blue triangles indicate joint upper limits obtained by combining prior observations with our follow-up observations, while grey triangles show limits derived solely from prior observations.
    Coincidentally, \citet{Good2023} conducted a follow-up observation of FRB~20190111A using Arecibo, which overlaps with one of our targets (red triangle). 
    The pink circle represents the repeating rate of FRB~20190110C, which was confirmed as a repeater by \citet{2023ApJ...947...83C} during the preparation of this manuscript. Its derived repeating rate is 0.95 hr$^{-1}$. We assume the properties of subsequent bursts are consistent with those of the initial burst, as detailed information was unavailable at the time of writing. 
    Error bars indicate the estimated confidence intervals: solid lines represent the 68\% intervals, while lighter, semi-transparent lines correspond to the 90\% intervals. Confidence intervals were calculated using the method of \citet{1991ApJ...374..344K}, which provides more accurate estimates when the Poisson event rate is low.
}
    \label{fig5}
\end{figure*}

\begin{figure*}
	\includegraphics[width=0.9\textwidth]{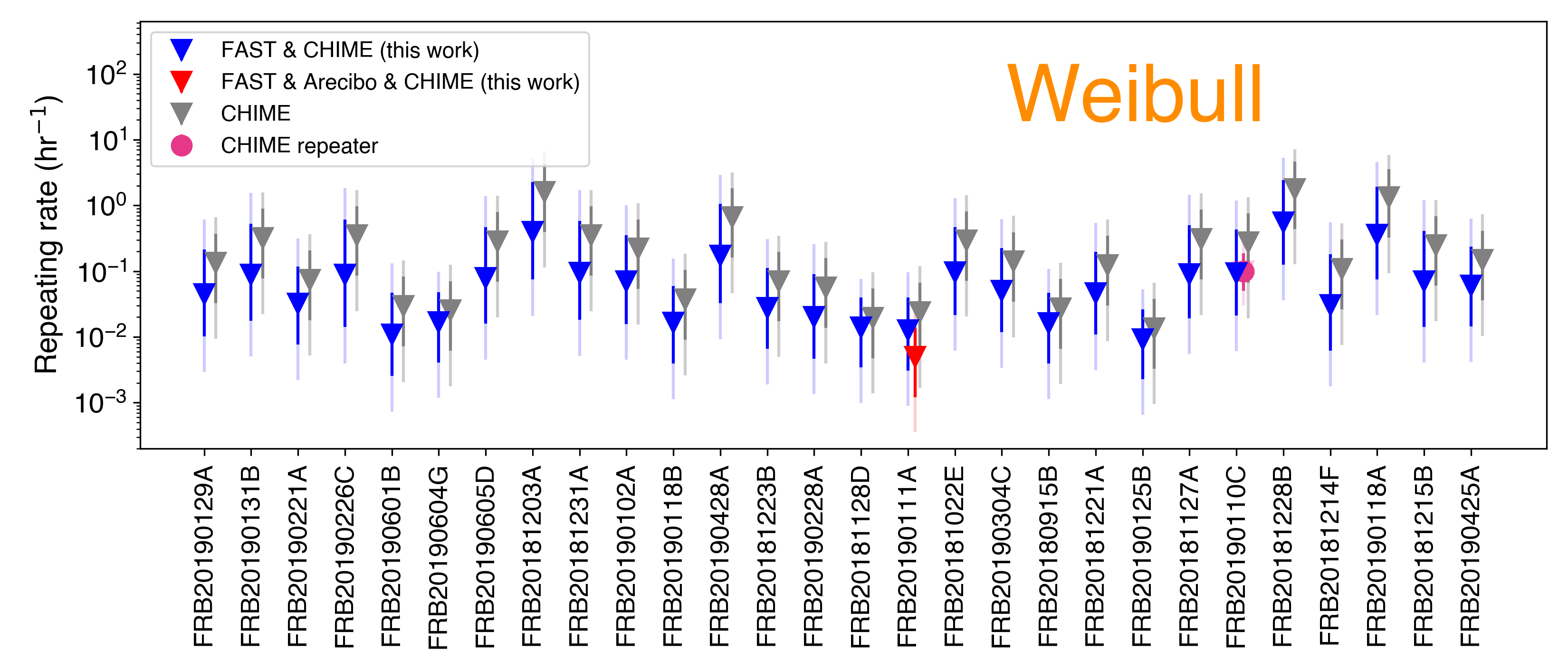}
    \caption{
   Upper limits on Weibull repeating rate ($r_{\rm scaled, joint}$) of 28 FRB sources derived by Eqs. (\ref{eq:r_scale_weibull0}) and (\ref{eq:r_scale_weibull1}). Only CHIME~FRB sources are included, with the duration of each daily observation, $\Delta$,  provided approximately by \citet{Good2023}, a necessary parameter for deriving the upper limit under the Weibull distribution.  The setup is otherwise identical to  Fig. \ref{fig5}, but with the Weibull model applied. The upper limit on repeating rate on the CHIME repeater FRB~20190110C is $0.142^{+0.106}_{-0.074}$ hr$^{-1}$. Error bars on each rate represent the uncertainties derived from the posterior probability distribution of $\log{r}$. The 50th percentile of the distribution is reported as the upper limit. Solid lines indicate 16th and 84th percentiles, corresponding to the 68\% confidence interval, while semi-transparent bars span the 5th to 95th percentiles, representing the 90\% confidence interval.
}
    \label{fig6}
\end{figure*}

Using the observational parameters
reported by \citet{2023ApJ...947...83C}, we calculate the repeating rate of the FRB~20190110C based on Eqs. (\ref{eq:r_scale}), (\ref{eq:r_scale_weibull0}), and (\ref{eq:r_scale_weibull1}).
The results are depicted as a magenta circle in Figs. \ref{fig5} and \ref{fig6} for both Poisson and Weibull repetition models.

In both Poisson and Weibull cases, the derived repeating rates are consistent with the upper limits placed by CHIME and FAST in this work within the errors.


\subsection{Positional errors}
FAST has 19 beams, each  with a size of approximately 0.05 degrees in diameter, resulting in a total field of view of about 0.4 degrees \citep[e.g.,][]{Jiang2020}.
This field of view is larger than the typical positional error of our targets selected from CHIME~FRB, which is around 0.2 degrees (see Tab. \ref{tab:frb-observations}). 
The specific beam sizes of other telescopes are as follows. 
The beam sizes of other telescopes used in FRB discoveries vary significantly. GBT discovered FRB~20110523A with a beam size of $\sim$ 0.3 degrees \citep{2015Natur.528..523M}.
Parkes,  through the High Time Resolution Universe (HTRU) survey, identified FRB~20110703A and FRB~20130729A, both with beam sizes of $\sim$0.2 degrees \citep{2013Sci...341...53T}.
Aditionally, the SUrvey for Pulsars and Extragalactic Radio Bursts (SUPERB) conducted with Parkes detected FRB~20151230A with a beam size of $\sim$0.1 degrees \citep{2018MNRAS.475.1427B}. 
Reanalysis of the Parkes Multibeam Pulsar Survey (PMPS) also identified FRB~20010621A, using a beam size of $\sim$0.1 degrees. 
UTMOST discovered FRB~20160410A with a significantly larger beam size of 4 degrees in the east-west direction, and 2.8 degrees in the north-south direction \citep{2017MNRAS.468.3746C}. 
Meanwhile, ASKAP, using the Commensal Real-time ASKAP Fast Transients (CRAFT), identified FRB~20171213A and FRB~20180212A, both with a beam size of $\sim$1 degree \citep{2017ApJ...841L..12B}. 

The positional errors and the sensitivity gaps between the 19 beams of FAST mentioned above may account for our non-detection of subsequent bursts in our repeating FRB candidates.
Therefore, we investigated the overlapped area of FAST 19-beams within positional uncertainty (see \ref{sub:upperlimit} for details). We found out our FAST observations cover about 10\% of positional uncertainty as shown in Tab. \ref{tab:result}. 

\section{conclusion}
\label{sec:conclusion}
Distinguishing between repeating and non-repeating FRB sources is essential for understanding the mysterious origin of FRBs, as these two classes may arise from fundamentally different progenitors.
However, identifying repeating sources requires logn-duration, high-sensitivity monitoring, which is observationally expensive and logistically challenging. 
To address this issue, we selected candidate repeaters among previously non-repeating FRBs using both empirical methods \citep{Hashimoto2019, Hashimoto2020, Kim2022}  and machine-learning techniques \citep{Chen2022}, and conducted follow-up observations with FAST.

We observed a total of 36 non-repeating FRB sources, each with an exposure time of 10 minutes.
Notably, one source selected by the machine-learning method was later confirmed by CHIME as a repeating FRB\citep{2023ApJ...947...83C}. Although the per-source exposure time was relatively short, FAST's exceptional sensitivity increases the likelihood of detecting faint bursts, especially under the assumption of a $-1.5$ slope in the FRB source count distribution, as reported by CHIME~FRBs \citep{CHIME2021_first}.

The FAST data were carefully inspected down to an S/N$\sim4$ after dedispersion using the nominal DM values of each source. No burst candidates exceeding S/N > 7 were identified, corresponding to a typical 7$\sigma$ fluence limit of approximately 0.013~Jy ms.
Our analysis improves the constraints on burst repetition rates by a factor of $\sim$ 3 compared to prior limits, suggesting a typical rate is lower than once per 20 hours, with possible indications of temporal clustering. 

This work establishes one of the stringent upper limits on FRB repeating rates to date, based on a smaple five times larger than that used in similar previous studies \citet{Good2023}.

\section*{Acknowledgements}
\label{sec:acknowledgements}
We express our sincere gratitude to Dr. Vishal Gajjar for 
serving as a referee and for his through review and insightful suggestions. His valuable feedback significantly contributed to improving the quality of this manuscript. 
We would also like  to extend our gratitude to Dr. Deborah C. Good for graciously taking the time to answer our questions.
YU is also grateful to Dr. Shotaro Yamasaki and Dr. Tomoki Wada for insightful discussions.
YU thanks the university for supporting research activities through the NCHU scholarship. 
TH acknowledges the support of the National Science and Technology Council of Taiwan through grants 110-2112-M-005-013-MY3, 110-2112-M-007-034-, 
111-2112-M-005-018-MY3,  
112-2123-M-001-004-, 
113-2112-M-005-009-MY3, 113-2123-M-001-008-,
 and Ministry of education of Taiwan trough the grant 113RD109. 
TG acknowledges the support of the National Science and Technology Council of Taiwan through grants 108-2628-M-007-004-MY3, 111-2112-M-007-021 and 112-2123-M-001-004-.
SH acknowledges the support of The Australian Research Council Centre of Excellence for Gravitational Wave Discovery (OzGrav) and the Australian Research Council Centre of Excellence for All Sky Astrophysics in 3 Dimensions (ASTRO 3D), through project number CE17010000 and CE170100013, respectively. JOC acknowledges financial support from the South African Department of Science and Innovation's National Research Foundation under the ISARP RADIOMAP Joint Research Scheme (DSI-NRF Grant Number 150551).
We acknowledge the use of the CHIME/FRB Public Database, provided at \url{https://www.chime-frb.ca/} by the CHIME/FRB Collaboration.
This research made use of Astropy, a community-developed core Python package for Astronomy \citep{Astropy2018}. 
This work has used the data from the Five-hundred-meter Aperture Spherical radio Telescope (FAST). 
FAST is a Chinese national mega-science facility, operated by the National Astronomical Observatories of Chinese Academy of Sciences (NAOC) \citep{2019SCPMA..6259502J, Jiang2020, 2020Innov...100053Q}.

\section*{Data Availability}
The data underlying this article has been released from FAST Operation and Development Center and is publicly available under the project ID: PT2021\_0076.



\bibliographystyle{mnras}
\bibliography{main} 




\appendix
\renewcommand{\thefigure}{A\arabic{figure}}
\begin{figure*}
	\includegraphics[width=0.8\textwidth]{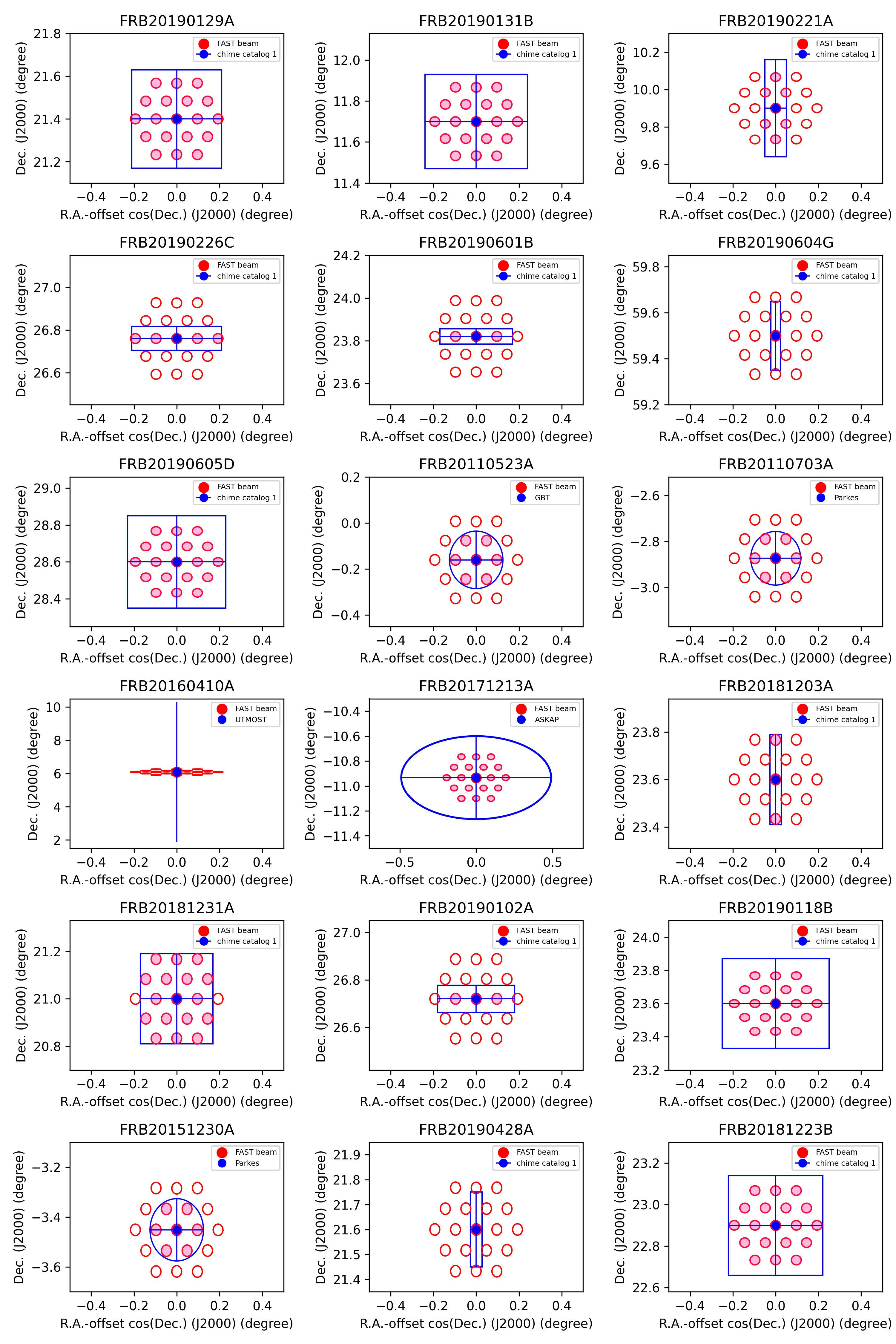}
    \caption{
    Positional errors of our FRB targets are shown here. The red circles represent the 19 beams of FAST, and the area marked in blue indicates the positional error region reported in the original discovery paper of FRBs \citep{CHIME2021_first, 2018Natur.562..386S}. 
    The region where the FAST beams overlap with the positional error area is coloured in pink. 
    Each figure employs a distinct y-axis scale, which reflects variations in the magnitude of the observed parameters due to different positional errors.
    }
    \label{fig:appendix_a1}
\end{figure*}

\begin{figure*}
	\includegraphics[width=0.8\textwidth]{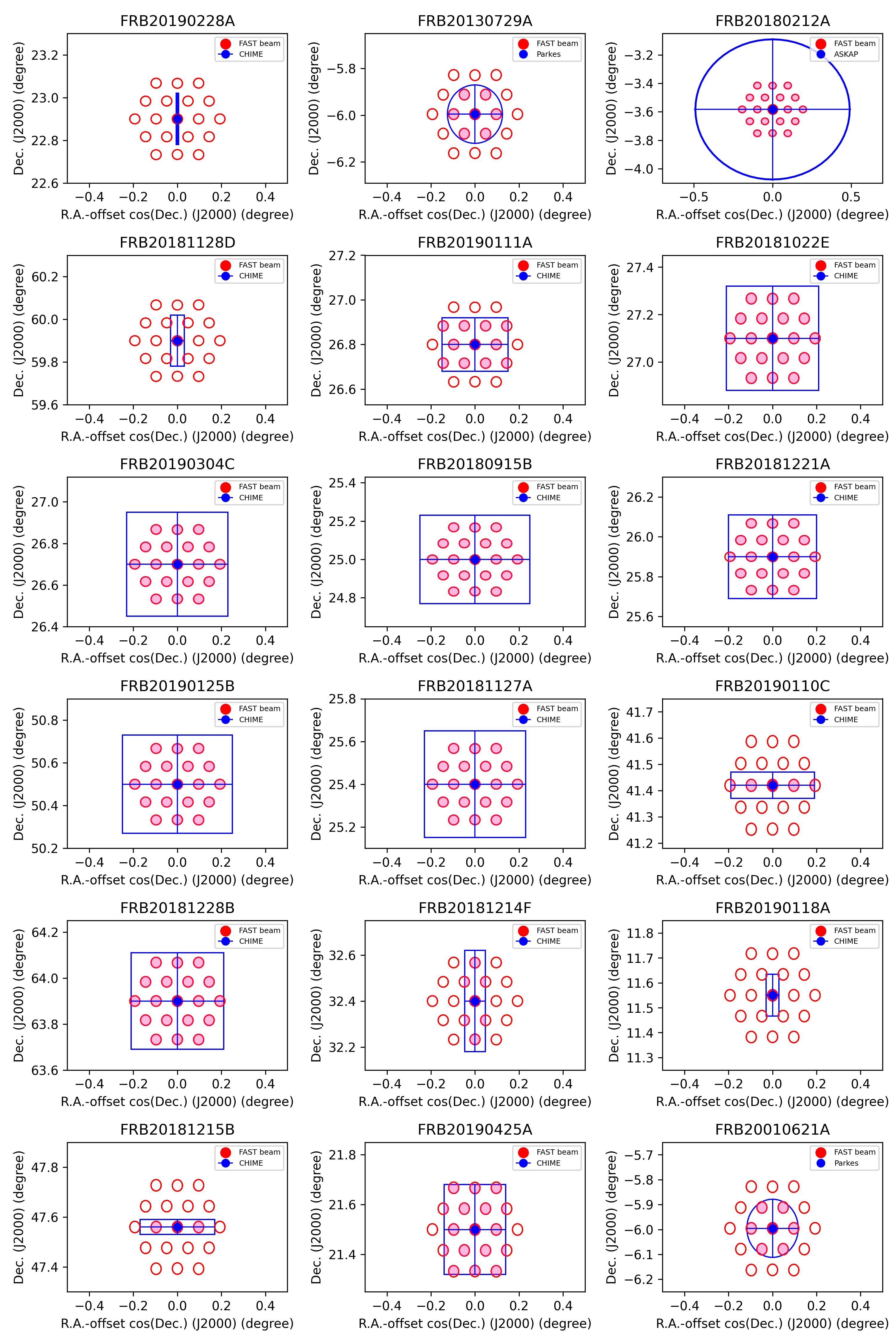}
    \caption{
    Figure \ref{fig:appendix_a2} is a continuation of Figure
    \ref{fig:appendix_a1}.
    }
    \label{fig:appendix_a2}
\end{figure*}

\section{Visualisation of the positional error regions}
\label{sec:appendix_A}
For a better understanding of the arguments about the chance of FRB source coverage with FAST (see Section \ref{sub:upperlimit}), we visualise the FAST beams and the positional error of our FRB targets from the original FRB detection papers \citep{CHIME2021_first, 2018Natur.562..386S}. 
Figs. \ref{fig:appendix_a1} and \ref{fig:appendix_a2} show FAST's 19 beams (red) and positional errors of our FRB targets (blue) in the sky. 
Each figure employs a distinct $y$-axis scale, which reflects variations in the positional errors of FRBs.

\renewcommand{\thefigure}{B\arabic{figure}}
\begin{figure*}
	\includegraphics[width=0.59\textwidth]{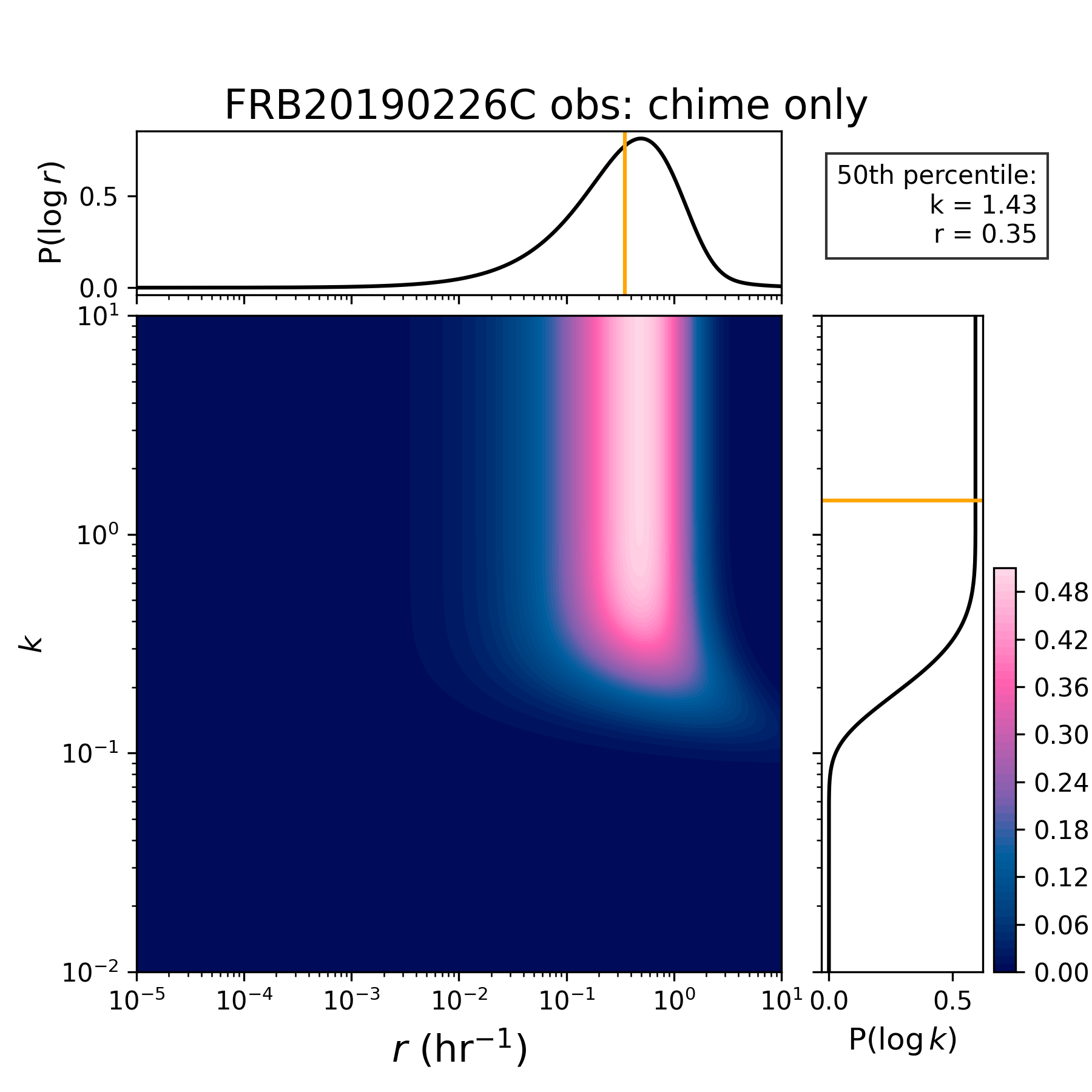}
    \caption{
    Posterior probability distribution of FRB~20190226C based on CHIME data only. The colour map shows the joint probability density, with pink indicating higher probability regions. The marginalized probability distributions for  $r$  and  $k$  are shown as black curves along the top and right axes, respectively. The orange line marks the 50th percentile of the marginalized distribution for $r$ and $k$.
    }
    \label{fig:appendix_b1}
\end{figure*}

\begin{figure*}
	\includegraphics[width=0.59\textwidth]{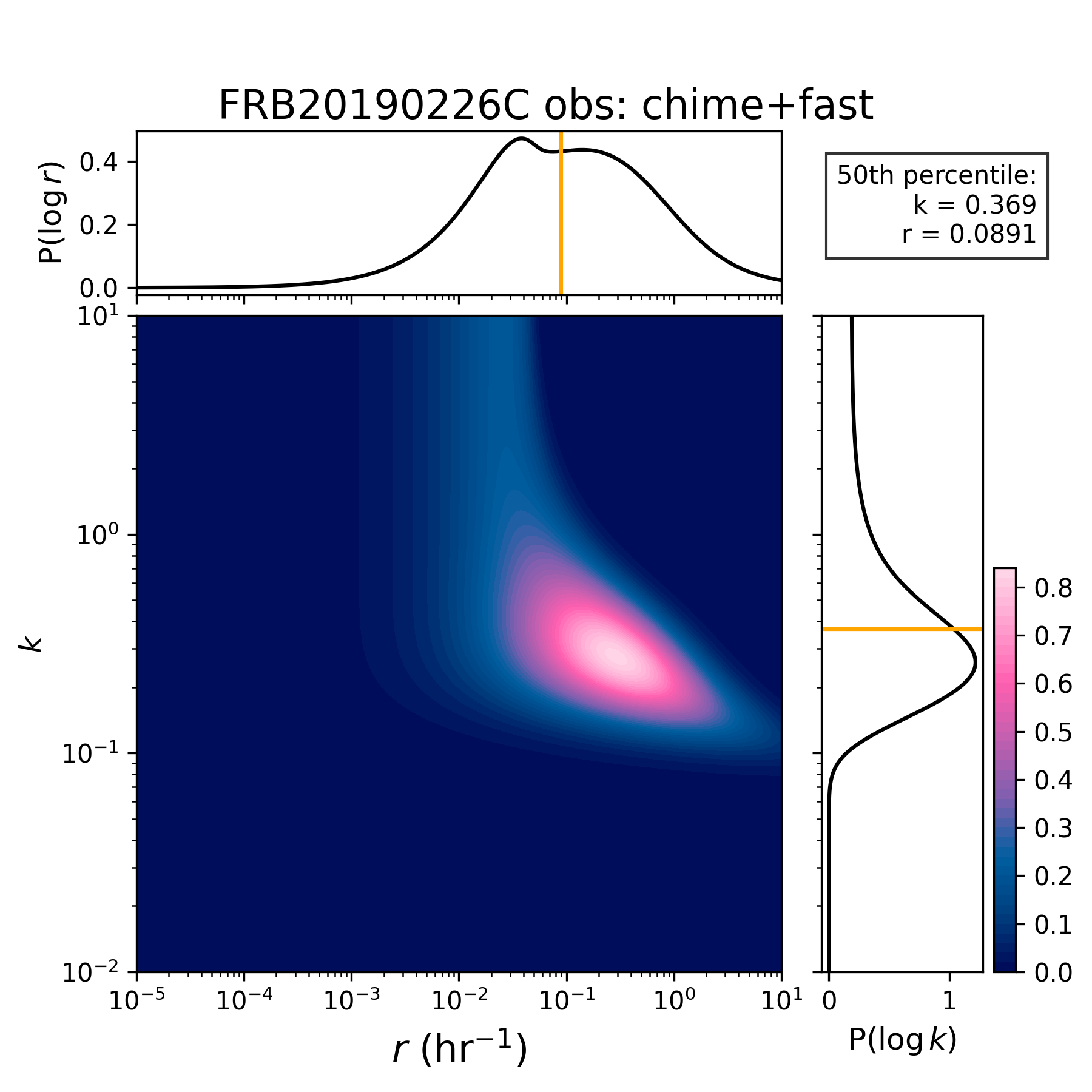}
    \caption{
    As in Figure \ref{fig:appendix_b1}, but based on the combined CHIME and FAST data for FRB~20190226C.
    }
    \label{fig:appendix_b2}
\end{figure*}

\begin{figure*}
	\includegraphics[width=0.59\textwidth]{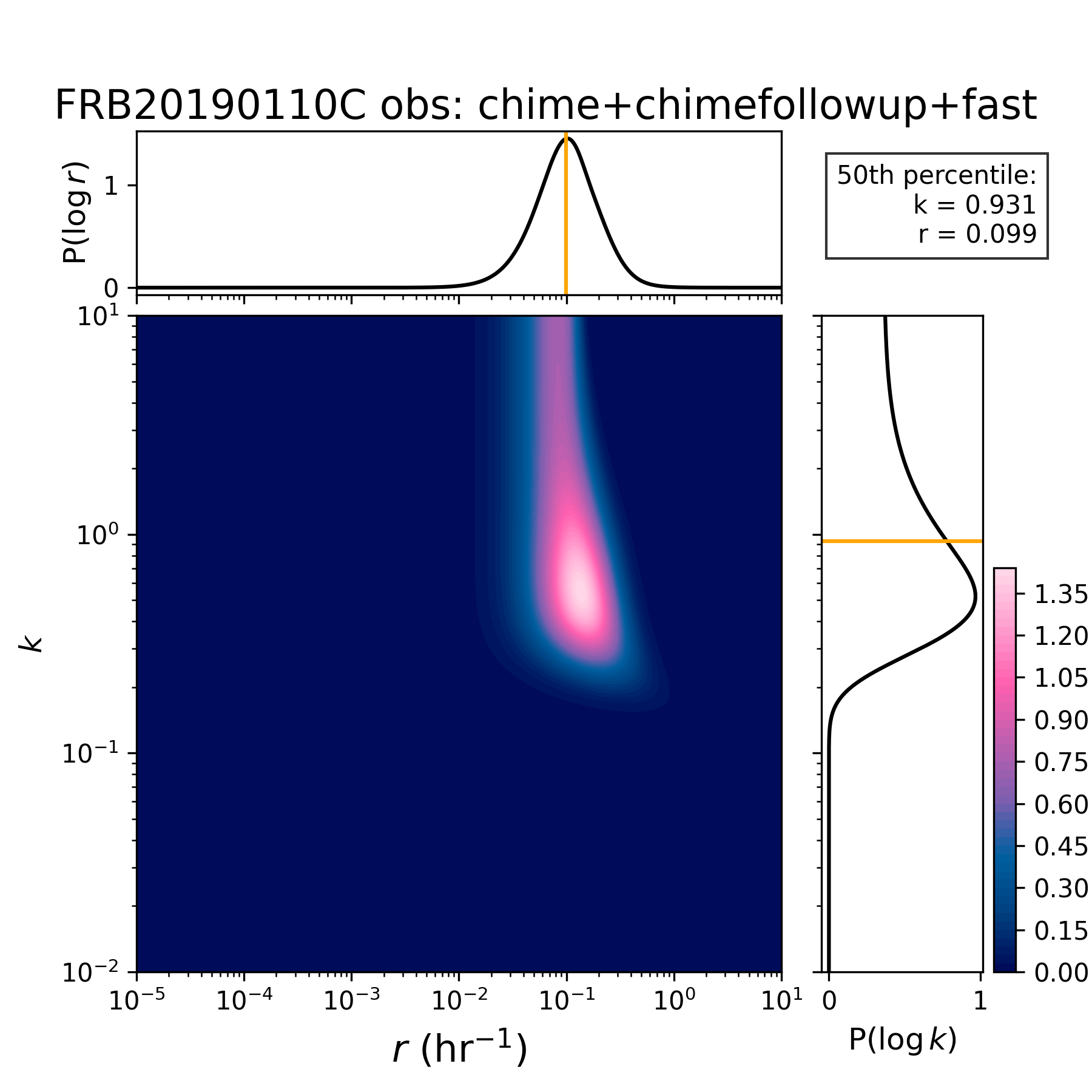}
    \caption{
    As in Figure \ref{fig:appendix_b1}, but for a different target, FRB~20190110C--confirmed as a repeater by CHIME--with combined CHIME, CHIME follow-up, and FAST data.
    }
    \label{fig:appendix_b3}
\end{figure*}

\section{Posterior probability on the Weibull model}
\label{sec:apppendix_b}
 The two-dimensional posterior PDFs of the event rate  $r$ and shape parameter $k$ under the Weibull model, along with their marginalized distributions in terms of $\log{r}$ and $\log{k}$, are presented in Figs. \ref{fig:appendix_b1}-\ref{fig:appendix_b3}.
Each figure highlights different aspects of the posterior distribution.

Fig. \ref{fig:appendix_b1} shows the posterir distribution for FRB~20190226C using CHIME data only.

Fig. \ref{fig:appendix_b2} presents the same source, but with combined CHIME and FAST data. The resulting distribution exhibits a somewhat bimodal structure, due to differences in sensitivity and observing time  between the two datasets.

Fig. \ref{fig:appendix_b3} shows the results for FRB~20190110C—a confirmed repeater—based on combined CHIME, CHIME follow-up, and FAST data. The posterior distributions for $\log{r}$ and $\log{k}$ appear more compact, reflecting the increased constraining power the CHIME follow-up observations, during which two additional bursts were detected \citet{2023ApJ...947...83C}. This is consistent with the general expectation that combining multiple detection datasets—analogous to multiplying Gaussian distributions—leads to tighter confidence intervals.



\bsp	
\label{lastpage}
\end{document}